\documentclass[superscriptaddress,amsmath,amssymb,longbibliography,aps]{revtex4-1}
\usepackage{graphicx}

\usepackage{mathrsfs}
\usepackage{mathbbol}
\usepackage{xcolor,booktabs}
\usepackage{overpic}
\usepackage{esint}
\usepackage{soul}

\usepackage[T1]{fontenc}
\usepackage{siunitx,mathrsfs}
\usepackage{multirow}
\usepackage{natbib} 
\usepackage{amsmath,amssymb}

\newcommand\ve[1]{\boldsymbol{#1}}

\renewcommand{\geq}{\geqslant}
\renewcommand{\leq}{\leqslant}
\newcommand{\upd}{\mathrm d}
\newcommand{\uppd}{\partial}
\newcommand{\const}{\text{const}}
\renewcommand{\mid}{\mathbin{|}}
\newcommand{\bx}{\boldsymbol x}

\newcommand{\m}{\text{m}}

\renewcommand{\v}{\text{v}}
\newcommand{\w}{\text{w}}
\renewcommand{\i}{\text{i}}
\renewcommand{\a}{\text{a}}
\renewcommand{\d}{\text{dry}}

\newcommand{\p}{{\text{p}}}
\newcommand{\Ddva}{D^{(2)}}
\newcommand{\Dodin}{D^{(1)}}

\newcommand{\dr}{\varepsilon_{s_\w}}
\newcommand{\K}{\text{K}}

\newcommand{\M}{\text{M}}

\newcommand{\Da}{\text{Da}}

\usepackage{overpic}
\graphicspath{{Figs/}}

\begin{document}
\title{Does small-scale turbulence matter for ice growth in mixed-phase clouds?}

\author{G. Sarnitsky}
\affiliation{Department of Physics, Gothenburg University, SE-41296 Gothenburg, Sweden} 
\author{G. Sardina}
\affiliation{Department of Mechanics and Maritime Sciences, Chalmers University of Technology, SE-41296 Gothenburg, Sweden}
\author{G. Svensson}
\affiliation{Department of Meteorology and Bolin Centre for Climate Research, Stockholm University, Stockholm, Sweden}
\affiliation{Department of Engineering Mechanics, KTH Royal Institute of Technology, Stockholm, Sweden}
\author{A. Pumir}
\affiliation{Univ. Lyon, ENS de Lyon, Univ. Claude Bernard, CNRS, Laboratoire de Physique, F-69342, Lyon, France}
\author{F. Hoffmann}
\affiliation{Meteorological Institute, Ludwig Maximilian University Munich, Theresienstr.\,37, 80333 Munich, German}
\author{B. Mehlig}
\email{Bernhard.Mehlig@gu.se}
\affiliation{Department of Physics, Gothenburg University, SE-41296 Gothenburg, Sweden} 

\begin{abstract}
Representing the glaciation of mixed-phase clouds in terms of the Wegener-Bergeron-Findeisen process is a challenge for many weather and climate models, which tend to overestimate this process because cloud dynamics and microphysics are not accurately represented. As turbulence is essential for the transport of water vapour from evaporating liquid droplets to ice crystals, we developed a statistical model using established closures to assess the role of small-scale turbulence. The model successfully captures results of direct numerical simulations, and we use it to assess the role of small-scale turbulence. We find that small-scale turbulence broadens the droplet-size distribution somewhat, but it does not significantly affect the glaciation time on submetre scales. However, our analysis indicates that  turbulence on larger spatial scales is likely to affect ice growth. While the model must be amended to describe larger scales, the present work facilitates a path forward to understanding the role of turbulence in the Wegener-Bergeron-Findeisen process.
\end{abstract}

\maketitle
\section{Introduction}
In mixed-phase clouds, ice particles grow at the cost of evaporating water droplets via the so-called Wegener-Bergeron-Findeisen (WBF) process \citep{storelvmo2015}. This occurs because the saturation vapour pressure differs for liquid water and ice. In the absence of any other processes, the WBF process turns mixed-phase clouds into ice clouds, which is commonly referred to as glaciation. However, mixed-phase clouds can be astonishingly stable (see e.g.\ \cite{morrison2012}), evading a too simplistic interpretation of the WBF process. Thus, our understanding and ability to adequately represent the WBF process has important implications on the longevity and coverage of mixed-phase clouds and hence Earth's radiation budget (e.g. Ref.~\cite{storelvmo2008}). 

\citet{korolev2008rates} used a rising-parcel model to study growth of ice particles  in mixed-phase clouds, under the assumption that ice particles, water droplets, water vapour and temperature are well mixed, showing that the cooling rates associated with a sufficiently strong updraft can prevent full glaciation. In a similar framework, \citet{ervens2011} highlighted the importance of the number of ice particle on the glaciation process, where smaller ice particle concentrations slow down glaciation. 

A question that has received little attention in connection with the WBF process is the impact of small-scale turbulence. 
Recently, \citet{chen2023mixed} used direct numerical simulations (DNSs) to study the influence of turbulence on growth of ice particles in a mixed-phase cloud, determining which conditions favour ice growth in cloud-top generating cells (CTGC).
The authors found that a higher liquid-water content (LWC) and higher relative humidity (RH) favour ice growth by the WBF process. 
Once the water droplets have evaporated, ice particles continue to grow consuming the remaining water vapour in the cloud. 
The simulations show that small-scale turbulence has a weak effect on the change of the mean radii of water droplets and ice particles, 
on the LWC and the ice-particle mass, and therefore on the glaciation time (defined as the time it takes to reach an ice-mass fraction of $0.9$). 
On the other hand, the simulations show how small-scale turbulence increases the width of the particle-size distributions.

\citet{chen2024pi} compiled parameter sets and specifications to compare models for mixed-phase processes in the Michigan Pi cloud chamber~\citep{chang2006laboratory}, including small-scale turbulence. The corresponding microscopic equations for the core region of the Pi chamber are similar to those of \citet{chen2023mixed}, except that a constant influx and removal of ice particles and droplets due to settling is specified.
 
Here, we analyse a statistical model for these processes, derived from the mapping-closure approximation \citep{pope1985pdf, chen1989probability, fries2024lagrangian} under the assumption that the Lagrangian supersaturation distributions (the distribution of supersaturation along droplet paths) are Gaussian, generalising statistical models for droplet-phase change in turbulence~\citep{kumala1997effect,siewert2017statistical,fries2021key,abade2017broadening,sardina2018broadening} to mixed-phase clouds. A strength of the model is that it is constructed from the microscopic governing equations. To assess the accuracy of the model, we compare its predictions to DNS results,
for the parameters of Ref.~\cite{chen2023mixed}, and for the Pi-chamber test case~\citep{chen2024pi}. Another advantage of the model is that it is straightforward to disregard small-scale turbulence in the model, simply by ignoring the stochastic terms. In this way the model simplifies to a parcel model for mixed-phase clouds \citep{korolev2003supersaturation,korolev2008rates,pinsky2018theoretical}. This allows us to study under which circumstances small-scale turbulence matters for glaciation, and when it does not. 
 Using the model we investigate the non-dimensional parameters of the problem, and discuss how our conclusions depend on the spatial scale of the turbulent fluctuations.

The remainder of this article is organised as follows. In Section \ref{sec:model} we describe the microscopic model for mixed-phase clouds, 
the basis for our DNS, and those of Refs.~\cite{chen2023mixed} and \cite{chen2024pi}. The statistical model is introduced in Section \ref{sec:sm}. Section \ref{sec:results} summarises our results, for  DNS, statistical model, and for its deterministic limit that disregards small-scale turbulence. In Section~\ref{sec:discussion} we compare the results and discuss their implications for glaciation of mixed-phase clouds. We summarise our conclusions in Section~\ref{sec:conclusions}. Three Appendices contain  a summary of all parameters used in the calculations, and mathematical details regarding our statistical-model-analysis.

\section{Microscopic model}
\label{sec:model}
\subsection{Supersaturation over ice and water}
\label{sec:supsat}
The microscopic model of \citet{chen2023mixed} describes how local temperature and the water-vapour mixing ratio are advected by the turbulent flow, and how droplets and ice particles grow and shrink in response to local fluctuations of these quantities. We start by showing how to simplify this microscopic dynamics by  combining the water-vapour and temperature fields into supersaturation fields over ice and water. Thereby we extend the results of Refs.~\cite{sardina2015continuous, fries2021key, pushenko2024connecting}
that treat the case without ice.  Water-vapour supersaturations $s_\w$ and $s_\i$ over liquid water and ice are defined via the partial pressure $p_\v$ of water vapour, and saturated vapour pressures $p_{\v, \w}$ and $p_{\v, \i}$ over liquid water and ice:
\begin{equation}\label{eq:s_p}
    s_\w = \frac{p_\v}{p_{\v, \w}} - 1, \qquad
    s_\i = \frac{p_\v}{p_{\v, \i}} - 1.
\end{equation}
For brevity,  we refer to ``liquid water'' as ``water'' and use subscripts ``$\w$'', ``$\i$'' and ``$\v$'' respectively for liquid water, ice and vapor. Expressions for $p_\w(T)$ and $p_\i(T)$ as functions of temperature $T$ are given by Eq.~\eqref{eq:pvs} in Appendix~\ref{app:parameters}. Supersaturations are often expressed in terms of the mixing ratio. The mixing ratio $q_\v$ of water vapour is defined as the ratio of the mass $m_\v$ of water vapour to the mass $m_\a$ of the dry air in a given volume, $q_\v = m_\v / m_\a$; or in terms of densities $q_\v = \rho_\v / \rho_\a$, where $\rho_\a= p_\a/({R_\a T})$ is the dry air density at partial air pressure $p_\a$. Note that the full pressure $p$ of the mixture is the sum of partial pressures of air and water vapour, $p = p_\v + p_\a$. But because $m_\v \ll m_\a$, we can take $\rho_\a \approx \rho$ (the density of the mixture) and $p_\a \approx p$, so that $p_\v = ({R_\v}/{R_\a}) p q_\v$, $R_\a$ and~$R_\v$ being the specific gas constants for dry air and water. Using this relation in Eqs.~\eqref{eq:s_p}, one can compute the supersaturation over water and ice as
\begin{equation}
  s_\w = \frac{R_\v}{R_\a} \frac{p}{p_{\v, \w}} q_\v - 1, \qquad
  s_\i = \frac{R_\v}{R_\a} \frac{p}{p_{\v, \i}} q_\v - 1 \label{eq:s}.
\end{equation}
In order to derive a consistent diffusion-convection-reaction equation for supersaturation, one approximates supersaturation as a linear function of $q_\v$, $T$ and $p$ near their reference values $q_{\v, 0}$, $T_0$ and $p_0$~\citep{fries2021key}. To this end we compute the differential of $s_\w$ from Eq.~\eqref{eq:s}:
\begin{align}
  \label{eq:full_ds}
     \upd s_\w
     &= (1 + s_\w) \biggl( \frac{\upd q_\v}{q_\v} - \frac{L_\w}{R_\v T} \frac{\upd T}{T} + \frac{\upd p}{p} \biggr) 
     = \frac{R_\v}{R_\a} \frac{p}{p_{\v, \w}} \upd q_\v+ (1 + s_\w) \biggl( - \frac{L_\w}{R_\v T} \frac{\upd T}{T} + \frac{\upd p}{p} \biggr),
\end{align}
where $L_\w$ is the latent heat of water evaporation,
and $L_\i$ is the latent heat of ice sublimation.
We determine $L_\w$ and $L_\i$ in the following way to ensure consistency with the approximations of $p_{\v,\w}$ and $p_{\v,\i}$,
\begin{equation}
    L_\w(T) = R_\v T^2 \frac{\upd \ln p_{\v, \w}}{\upd T}, \quad
    L_\i(T) = R_\v T^2 \frac{\upd \ln p_{\v, \i}}{\upd T}.
\end{equation}
We can further simplify Eq.~\eqref{eq:full_ds}. Within the Oberbeck--Boussinesq approximation, the variations of $p$ and~$T$ around $p_0$ and~$T_0$ are small, allowing us to use constant coefficients in front of the differentials. Next, we deal with the factor $1 + s_\w$. We assume that the supersaturation variations $\Delta s_\w$ satisfy $\Delta s_\w \ll 1 + s_{\w, 0}$, where
\begin{equation}\label{eq:s_w_0}
  s_{\w, 0} = s_\w(q_{\v, 0}, T_0, p_0)
  =\frac{R_\v}{R_\a} \frac{p_0}{p_{\v, \w}(T_0)} q_{\v, 0} - 1.
\end{equation}
We stress that this assumption is violated when the variations of $s_\w$ (equivalently of $q_\v$) are not small, as
for example when completely dry air at $s_\w = -1$ saturates to~\hbox{$s_\w = 0$}. 

When the supersaturation fluctuations are small enough, we can integrate Eq.~\eqref{eq:full_ds} using the simplifying assumptions of the previous paragraph
to obtain $s_\w$ as a linear function of $q_\v$, $T$ and~$p$ (and $s_\i$ is derived in a similar manner):
\begin{subequations}\label{eq:linear_s}
    \begin{align}\label{eq:linear_s_w}
      s_\w  &= s_{\w,0} + \frac{R_\v}{R_\a} \frac{p_0}{p_{\v, \w}(T_0)} (q_\v - q_{\v,0})  + (1 + s_{\w, 0}) \biggl(- \frac{L_\w(T_0)}{R_\v T_0} \frac{T - T_0}{T_0} + \frac{p - p_0}{p_0} \biggr), \\
    \label{eq:linear_s_i}
      s_\i  &= s_{\i,0} + \frac{R_\v}{R_\a} \frac{p_0}{p_{\v, \i}(T_0)} (q_\v - q_{\v,0}) + (1 + s_{\i, 0}) \biggl( - \frac{L_\i(T_0)}{R_\v T_0} \frac{T - T_0}{T_0} + \frac{p - p_0}{p_0} \biggr).
    \end{align}
\end{subequations}
In order to derive the diffusion-convection-reaction equations for the fields $s_\w(\bx, t)$ and $s_\i (\bx, t)$ from Eqs.~(\ref{eq:linear_s}), we follow~\cite{sardina2015continuous} and start from the corresponding equations for the fields $T(\bx, t)$ and  $q_\v(\bx, t)$:
\begin{subequations}
\label{eq:dynTq}
\begin{align}
  \frac{\upd T}{\upd t}
  &= 
   \varkappa_T \frac{\uppd^2 T}{\uppd x_j \uppd x_j} + \frac{L_\w(T_0)}{c_p} C_\w + \frac{L_\i(T_0)}{c_p} C_\i, \label{eq:T} \\[1ex]
  \frac{\upd q_\v}{\upd t}
  &= \varkappa_{q_\v} \frac{\uppd^2 q_\v}{\uppd x_j \uppd x_j} - C_\w - C_\i. \label{eq:qv}
\end{align}
\end{subequations}
Here $C_\w(\bx, t)$ and $C_\i(\bx, t)$ are water and ice condensation and deposition rates, which are discussed in more detail in Section~\ref{sec:condensation_rates}, $c_p = \frac{7}{2} R_\a$ is the specific heat of air at constant pressure, $\varkappa_T$ and $\varkappa_{q_\v}$ are the molecular diffusivities of~$T$ and $q_\v$. Math-style Latin indices (as in $x_j$) denote vector/tensor components in Cartesian coordinates, and we use the Einstein summation convention for repeated indices. Next, $\upd / \upd t = \uppd /\uppd t + u_j \uppd / \uppd x_j$ are the components of the convective derivative, where $u_j(\bx, t)$ is the turbulent velocity of air determined by the Navier--Stokes equations
\begin{equation}
\label{eq:ns}
  \frac{\upd u_j}{\upd t} = -\frac{1}{\rho_0} \frac{\uppd p}{\uppd x_j} + \nu \frac{\uppd^2 u_j}{\uppd x_k \uppd x_k} + f_{j}^{(u)}, \qquad \frac{\uppd u_j}{\uppd x_j} = 0,
\end{equation}
Here $\nu$ is the kinematic viscosity of air. The forcing $f_{j}^{(u)}$ is  required to maintain stationary turbulence, as in Refs.~\cite{chen2023mixed,chen2024pi}. 

Now we combine Eqs.~(\ref{eq:linear_s}) and (\ref{eq:dynTq}). For this, we first calculate the Laplacian of $s_\w$ from Eq.~(\ref{eq:linear_s_w}). Within the Oberbeck--Boussinesq approximation only the hydrostatic pressure affects the thermodynamic variables like~$s_\w$, and the Laplacian of hydrostatic pressure is negligible compared to the effects of Laplacians of~$T$ and~$q_\v$, which are dominated by small-scale turbulence. Thus, we can neglect the Laplacian $\uppd^2 p / {\uppd x_j \uppd x_j}$ to obtain
\begin{align}
  \frac{\uppd^2 s_\w}{\uppd x_j \uppd x_j}& = \frac{R_\v}{R_\a} \frac{p_0}{p_{\v,\w}(T_0)} \frac{\uppd^2 q_\v}{\uppd x_j \uppd x_j}
  - \frac{(1 {+} s_{\w, 0}) L_\w(T_0)}{R_\v T_0^2} \frac{\uppd^2 T}{\uppd x_j \uppd x_j}.
\end{align}
The second step is to assume $ \varkappa_{q_\v} \approx \varkappa_{T}$, which is justified since $\varkappa_{q_\v}  = 1.17 \varkappa_{T}$ (Eq.~\eqref{eq:vkk}). Defining $\varkappa = \sqrt{\varkappa_{q_\v} \varkappa_T}$ allows us to write
\begin{subequations}\label{eq:2_s}
\begin{align}
  \frac{\upd s_\w}{\upd t}
  &= \varkappa \frac{\uppd^2 s_\w}{\uppd x_j \uppd x_j}  - A_{2, \w, \w} \, C_\w - A_{2, \w, \i} \, C_\i, \label{eq:6A_s_w} \\
  \frac{\upd s_\i}{\upd t}
  &= \varkappa \frac{\uppd^2 s_\i}{\uppd x_j \uppd x_j} - A_{2, \i, \w} \, C_\w - A_{2, \i, \i} \, C_\i \label{eq:6A_s_i},
  \end{align}
\end{subequations} 
where the equation for $s_\i$ is derived analogously. Here  we introduced the parameters
\begin{align}
 \label{eq:6A_A_2}
  A_{2, \phi_1, \phi_2}
 & = \frac{R_\v}{R_\a} \frac{p_0}{p_{\v, \phi_1}(T_0)}
 + \frac{(1 + s_{\phi_1, 0}) L_{\phi_1}(T_0) L_{\phi_2}(T_0)}{c_p R_\v T_0^2}.
\end{align}
In this expression, the variable $\phi$ stands for a particular condensed phase, either $\phi = \w$ or $\phi = \i$. If we disregard the ice phase, we obtain the supersaturation dynamics used in the statistical models for evaporation of water droplets at the cloud edge~\citep{fries2021key,fries2024lagrangian}. There are minor differences in the expressions of the $A$-parameters in those references, reflecting slightly different assumptions. 

The model~\eqref{eq:2_s} can be further simplified if we express~$s_\i$ as a  
function of $s_\w$, so that it is sufficient to solve a single partial differential equation for~$s_\w$. The final model for $s_\w$ and $s_\i$ used in our DNS and in the statistical model is
\begin{subequations}\label{eq:DNS}
\label{eq:supsat} 
\begin{align}
  \frac{\upd s_\w}{\upd t}
  &= \varkappa \frac{\uppd^2 s_\w}{\uppd x_j \uppd x_j} - A_{2, \w} C_\w - A_{2, \i} C_\i + f^{({s_\w})}, \label{eq:DNS:s_w} \\
  s_\i
  &= A_4 (s_\w + 1) - 1.\label{eq:DNS:s_i} 
\end{align}
\end{subequations}
Here we introduced a forcing term $f^{(s_\w)}$ representing  the forcing of small-scale fluctuations due to large spatial scales that are not resolved by the DNS~\citep{chen2023mixed,chen2024pi}. The new $A_2$-parameters are given by $A_{2, \w} = A_{2, \w, \w}$ and $A_{2, \i} = A_{2, \w, \i}$, in simplified notation. The parameter $A_4$ is defined as:
\begin{equation}
    A_4 = \frac{p_{\v, \w}(T_0)}{p_{\v, \i}(T_0)}
\end{equation}
Since $p_{\v, \i} > p_{\v, \w}$ for $T < \qty{0}{\celsius}$, we have $A_4 > 1$ and $s_\i > s_\w$ in the mixed-phase cloud. The WBF process corresponds to $s_\i > 0 > s_\w$, under this condition water droplets evaporate and ice particles grow. To assess how the single\--super\-saturation approximation works, we estimate the error $\delta s_\i$ between $s_\i$ calculated from~\eqref{eq:linear_s_i} and~\eqref{eq:DNS:s_i}. For $T_0$ between 
$\qty{231.15}{K}$ and $ \qty{273.15}{K}$ and the parameters 
from Appendix~\ref{app:parameters} we find:
\begin{equation}
   \delta s_\i/s_\i
    \approx -\qty{0.011}{K^{-1}} \,(T - T_0) / s_\i .
\end{equation}
We conclude that the single-supersaturation
approximation~\eqref{eq:DNS:s_i} works well for temperature variations of the order of $T - T_0 \sim \qty{1}{K}$, provided that $|s_\i| \gg 0.01$. 
The latter condition is satisfied at the initial and most interesting stage of glaciation, when air is saturated with respect to water, $s_\w =0$ and $s_\i > 0.1$. More generally, the approximation works well when temperature fluctuations are much smaller than $\qty{1}{K}$, as in the case of the CTGC~\cite{chen2023mixed}.

\subsection{Particle dynamics}
\label{sec:pdyn}
Water droplets and ice particles are assumed to be so small that they follow the flow, their positions $\bx_\w(t)$ and $\bx_\i(t)$ obey
\begin{align}
\label{eq:advection} 
  \frac{\upd x_{\w, j}}{\upd t} = u_j(\bx_{\w}, t), \qquad
  \frac{\upd x_{\i, j}}{\upd t} = u_j(\bx_{\i}, t).
\end{align}
In other words, effects of particle inertia \citep{bec2024statistical} are neglected.
The particle  radii $r_\w(t)$ and $r_\i(t)$  change according~to:
\begin{subequations}\label{eq:DNS:r}
    \begin{align}
    \label{eq:dnsr1}
        \frac{\upd r_\w^2}{\upd t}
        &= 2 A_{3, \w} \, a_3(r_\w / r_{A_{3, \w}}) \, \bigl[ s_\w - s_{\w, \K}(r_\w) \bigr], \\
        \frac{\upd r_\i^2}{\upd t} \label{eq:DNS:r_i}
        &= \biggl\{\begin{aligned}
            &2 A_{3, \i} \, a_3(r_\i / r_{A_{3, \i}}) \, s_\i &\text{if } r_\i > 0, \\
            &0 &\text{if } r_\i = 0,
        \end{aligned}
    \end{align}
\end{subequations}
with supersaturation taken at the particle position, e.g.\ $s_\w(t) = s_\w(\bx_\w, t)$. For ice particles, 
${\rm d}r_\i/{\rm d}t$ is constrained to vanish at $r_\i=0$ to ensure that $r_\i^2$ remains non-negative, as it must. 

Water droplets are not allowed to completely evaporate, due to
the K\"ohler correction term in Eq.~(\ref{eq:DNS:r}), involving the radius-dependent function $s_{\w, \K}$. This function is parameterized by the dry aerosol radius $r_\d$ and the hygroscopicity coefficient~$\kappa$:
\begin{align}
\label{eq:swK}
  s_{\w, \K}(r_\w)  &= \frac{r_\w^3 - r_\d^3}{r_\w^3 - r_\d^3 (1 - \kappa)} - 1.
\end{align}
A more general expression for K\"ohler corrections contains an exponential term for 
 Kelvin curvature effects~\citep{Petters_2007}. In Eq.~(\ref{eq:swK}) we approximated the exponential by unity.
This is a good approximation for our values of $r_\d$. 
Unlike water droplets, ice particles are not allowed to reactivate: once an ice particle evaporates and~$r_\i$ reaches zero, it stays evaporated with $r_\i = 0$ (the ice is pure). This is in agreement with the specifications of both the CTGC and the Pi Chamber cases. Equations~\eqref{eq:DNS:r} contain corrections for the efficiency of accommodation of water vapour on the particle surface, introducing the particle-size dependent function $a_3(x) = x/(x + 1 )$ with accommodation length $r_{A_{3, \phi}}$ \citep{kogan1991simulation, jeffery2007inhomogeneous}:
\begin{equation}\label{eq:r_A_3}
 r_{A_{3, \phi}} = A_{3, \phi} \frac{\rho_\phi \sqrt{2 \pi R_\v T_0}}{\alpha_{q_\v, \phi} \, p_{\v, \phi}(T_0)},
\end{equation}
where $\phi$ refers to the phase of condensed water, either $\phi = \w$ or $\phi = \i$. Particles with radii $r_\phi \gg r_{A_{3, \phi}}$ are not affected by these corrections. Here $\alpha_{q_\v, \w}$ and $\alpha_{q_\v, \i}$ are water-vapour accommodation coefficients over water and ice. The use of the particular forms of $A_{3, \phi}$, $a_3$ and $r_{A_3, \phi}$ for our cases is justified in Appendix~\ref{app:parameters}.  Neglecting all radius-dependent corrections corresponds to $a_3 = 1$ and $s_{\w, K} = 0$. The $A_3$-parameters in Eqs.~(\ref{eq:DNS:r}) and (\ref{eq:r_A_3}) are given by \citep{Mordy_1959} 
\begin{equation}\label{eq:A_3}
  A_{3, \phi} = \biggl[ \frac{R_\a}{R_\v} \frac{\rho_\phi L_\phi^2(T_0)}{\varkappa_T c_p T_0 p_0} + \frac{\rho_\phi R_\v T_0}{\varkappa_{q_\v} p_{\v, \phi}(T_0)}\biggr]^{-1}.
\end{equation}
Finally, the specifications for the cloud-chamber test case allow for injection and removal of water droplets and ice particles \citep{chen2024pi}. Water droplets are injected  in the form of dry aerosol  at a constant rate $I_\w \, [(\text{s} \, \m^3)^{-1}]$ and removed at a rate defined by the settling velocity $u_{\infty, \w}$. Ice particles are treated the same way, with $I_\i$ and~$u_{\infty, \i}$. Namely, each particle is removed with a probability $P_\w$ (droplet) or $P_\i$ (ice particle):
\begin{equation}
  P_{\w} = \min \Bigl( \frac{u_{\w, \infty} \Delta t}{H} , 1 \Bigr), \quad
  P_{\i} = \min \Bigl( \frac{u_{\i, \infty} \Delta t}{H} , 1 \Bigr)
\end{equation}
where $H$ is is the total height of the Pi chamber. The settling velocities depend on the particle radii $r_\w$ and $r_\i$:
\begin{align}
\label{eq:setvel}
  u_{\infty, \w} &= k_{\infty, \w} r_\w^2, \qquad u_{\infty,\i} = k_{\infty, \i} r_\i^2.
\end{align}
The values of the parameters $k_{\infty, \w}$ and~$k_{\infty, \w}$ are specified by~\cite{chen2024pi}. We note that particle shape  impacts the sedimentation velocity. Here we assume spherical particles, but larger ice crystals (with radii > 30$\mu$m)
tend to be non-spherical. This is not accounted for in the model (for the data discussed below, Figures~\ref{fig:1} and \ref{fig:2}, the ice particles do not exceed this size).
The overall number of particles changes as
\begin{equation}\label{eq:pn}
  \frac{\upd N_\w}{\upd t} = V I_\w + \frac{N_\w}{H} \langle  u_{\w, \infty} \rangle, 
  \frac{\upd N_\i}{\upd t} = V I_\i + \frac{N_\i}{H}\langle u_{\i, \infty} \rangle,
\end{equation}
where $V$ is the volume of the simulation domain (corresponding to the core of the Pi chamber).

\subsection{Condensation and deposition rates}\label{sec:condensation_rates}
The condensation and deposition rates $ C_\w$, $C_\i$ for water and ice are defined through the rate of change of condensed water content:
\begin{subequations}\label{eq:C_def}
\begin{align}
  C_\w(\bx, t) \label{eq:Cw_def}
  &= \frac{4}{3} \pi \frac{\rho_\w}{\rho_0} \sum_{\alpha=1}^{N_\w} G(\bx - \bx_{\w,\alpha}) \frac{\upd r_{\w, \alpha}^3}{\upd t}, \\
  C_\i(\bx, t) \label{eq:Ci_def}
  &= \frac{4}{3} \pi \frac{\rho_\i}{\rho_0} \sum_{\alpha=1}^{N_\i} G(\bx - \bx_{\i,\alpha}) \frac{\upd r_{\i, \alpha}^3}{\upd t},
\end{align}
\end{subequations}
where $N_\w$ is the number of water droplets,  $N_\i$ is the number of ice particles, and $G$ is the standard spatial kernel, normalized to unity~\citep{vaillancourt2002microscopic,chen2023mixed}. The spatial range of $G$ is  the linear size of a DNS-grid cell. Using Eqs.~\eqref{eq:DNS:r} we can rewrite $C_\w$ and $C_\i$ as
\begin{subequations}\label{eq:condrates}
\begin{align}
  C_\w(\bx, t)
  &= 4 \pi \frac{\rho_\w}{\rho_0} A_{3, \w} \sum_{\alpha=1}^{N_\w} G(\bx - \bx_{\w,\alpha}) r_{\w, \alpha} a_3(\tfrac{r_{\w, \alpha} }{r_{A_{3, \w}}}) \bigl[ s_{\w, \alpha} - s_{\w, \K}(r_{\w, \alpha}) \bigr],\\
  C_\i(\bx, t)
  & =  4 \pi \frac{\rho_\i}{\rho_0} A_{3, \i}  \sum_{\alpha=1}^{N_\i}\! G(\bx - \bx_{\i,\alpha}) r_{\i, \alpha} a_3(\tfrac{r_{\i, \alpha} }{r_{A_{3,\i}}}) s_{\i, \alpha}.\label{eq:condrates_i}
\end{align}
\end{subequations}
Here $s_{\w, \alpha}$ is the supersaturation field at the position of particle $\alpha$, $s_{\w, \alpha}(t) = s_\w(\bx_{\alpha}, t)$. Note that the multiplication by $r_{\i, \alpha}$ in Eq.~\eqref{eq:condrates_i} correctly accounts for the $r_\i = 0$ condition in Eq.~\eqref{eq:DNS:r_i}.

Without injection and sedimentation ($N_\w$ and $N_\i$ are constant) and for vanishing mean forcing $\langle f^{(s_\w)} \rangle$, Eqs.~\eqref{eq:DNS:s_w} and \eqref{eq:C_def} imply
that  the quantities 
\begin{subequations}\label{eq:s_inv}
    \begin{align}
        s_{\w, \text{inv}} &=
        \langle s_\w \rangle_V + \frac{4}{3} \pi \frac{\rho_\w}{\rho_0} A_{2, \w} n_\w \langle r_\w^3 \rangle_V + \frac{4}{3} \pi \frac{\rho_\i}{\rho_0} A_{2, \i} n_\i \langle r_\i^3 \rangle_V, \\
        s_{\i, \text{inv}} &= A_4(s_{\w, \text{inv}} + 1) - 1. \label{eq:s_i_inv}
    \end{align}
\end{subequations}
are invariant. 
Here $n_\w = N_\w / V$ and $n_\i = N_\i / V$ are droplet and ice number densities, and $\langle s_\w \rangle_V = \frac{1}{V}\int s_\w \, \upd\bx$ is the spatially averaged supersaturation. Since Eq.~\eqref{eq:DNS:s_w} for~$s_\w$ is derived from the two Eqs.~\eqref{eq:dynTq} for $q_\v$ and for $T$, the conservation law~\eqref{eq:s_inv} combines both water and thermal energy conservation during condensation and deposition of water vapor~\citep{siewert2017statistical, fries2021key}.
Physically, $s_{\w, \text{inv}}$ and $s_{\i, \text{inv}}$ correspond to supersaturation over water and ice if all the particles evaporate, so that all the water is contained in the form of water vapor. In case we ignore the K\"ohler corrections, $s_{\i, \text{inv}}$ provides us with an insight into the final state of the cloud. Since water evaporates ($r_\w = 0$), $s_{\i, \text{inv}} > 0$ means that ice remains, while  $s_{\i, \text{inv}} \leq 0$ implies that ice evaporates too.

All thermodynamic parameters in the above microscopic equations are summarised in 
Tables~\ref{tab:A} (CTGC) and \ref{tab:B} (Pi chamber) in  Appendix~\ref{app:parameters}.

\subsection{Direct numerical simulations}
\label{sec:dns}
We performed DNSs using Eqs.~(\ref{eq:ns}) to (\ref{eq:pn}) for mixed-phase processes in the core of the Pi chamber, for the parameters specified by~\citet{chen2024pi}. The turbulent dissipation 
rate per unit mass was $\varepsilon=66$ ${\rm cm}^2{\rm s}^{-3}$ in a cubic domain 
with side length $L=20$ cm. With $\nu = 1.278\times 10^{-5}$  m$^2$s$^{-1}$, the Kolmogorov length $\eta =(\nu^3/\varepsilon)^{1/4}$ is around $\eta=0.75$ mm. To properly resolve the turbulent flow we used a numerical resolution with $256^3$ collocation points to solve the Eulerian equations. 

As specified by~\citet{chen2024pi}, our DNSs used the forcing term in the Navier--Stokes equations (\ref{eq:ns}) to maintain  a statistically steady turbulent state with constant dissipation rate $\varepsilon$. In Fourier space, the forcing reads 
\begin{align}
     \hat {\mbox{}\!\ve f}^{(u)}(\ve k)=\varepsilon  \mathcal{N}(t)  
    \hat{\ve u}
    (\ve k,t)\quad\mbox{for $|\ve k| < 3\pi/L$}.
\end{align}
Here  $\hat {\ve u}({\ve {k}},t)$ is the Fourier transform of the turbulent velocity field $\ve u(\ve x,t)$, $\ve k$ is the wave vector, and  ${\mathcal N}(t)  = 
\big(\sum_{|\ve k| < 2\pi/L} {\ve {|\hat u}}({\ve {k}},t)|^2\big)^{-1}$ is a normalisation factor.
The supersaturation equation (\ref{eq:DNS:s_w}) is also forced. For the CTGC we use a Gaussian random forcing~\citep{paoli2009turbulent}, 
\begin{align}
\label{eq:ssf}
    \hat{f}^{(s_\w)}(\ve k)=\beta {\rm d}W(\ve k,t)\quad\mbox{for $|\ve k| < 3\pi/L$},
\end{align}
where ${\rm d}W(t)$ is white noise with  unit variance independently chosen for different $\ve k$. 
The numerical factor $\beta$  is used  to maintain the prescribed steady water supersaturation root mean square $\sigma_{s_\w}$ before the aerosol injection. The  forcing (\ref{eq:ssf})  differs from the one suggested by \citet{chen2024pi} where there is no explicit forcing term, instead the supersaturation Fourier coefficients of the forced wavenumbers are rescaled at each time step to mantain the prescribed $\sigma_{s_\w}$. We tested that the two forcing schemes yield the same condensation/evaporation statistics, provided  that they achieve the same statistically steady-state  value of~$\sigma_{s_\w}$. For the Pi chamber, \citet{chen2024pi} specify  an additional average forcing in Eq.~(\ref{eq:DNS:s_w}) that nudges the average supersaturation,
\begin{equation}
\label{eq:fs}
    \langle f^{({s_\w})} \rangle = -  \bigl( \langle s_\w \rangle - s_{\w, \text{force}} \bigr)/{\tau_{s_{\w, \text{force}}}},
\end{equation}
where $s_{\w,{\rm force}}$ is the mean supersaturation before aerosol injection, and $\tau_{s_{\w, \text{force}}}$ is a forcing timescale. This forcing mimics the property of the Rayleigh–Benard convection inside the cloud chamber to achieve a statistically steady thermodynamic state \citep{saito19}.

For the Pi chamber,  we need to add and remove the particles during each time step~$\Delta t$ as specified by~\citet{chen2024pi}, see Eqs.~(\ref{eq:setvel}) and (\ref{eq:pn}) in Section \ref{sec:pdyn}.  To this end, the numbers of added particles with radii $r_{\w, \text{initial}}$ and $r_{\i, \text{initial}}$ are
\begin{align}
  \Delta N_{\w} &= \mathrm{floor}(I_\w V\Delta t) +
  \biggl\{ \begin{aligned}
    &1, && \mathrm{frac}(I_\w V\Delta t) \geq \xi_\w, \\
    &0, && \text{else}, \\
  \end{aligned}\\
  \Delta N_{\i} &= \mathrm{floor}(I_\i V\Delta t) +
  \biggl\{ \begin{aligned}
    &1, && \mathrm{frac}(I_\i V\Delta t) \geq \xi_\i, \\
    &0, && \text{else},
  \end{aligned}
\end{align}
where $I_\w$ and $I_\i$ are water and ice injection rates, and $\xi_\w$ and $\xi_\i$ are independent random variables uniformly distributed in [0,1]. The initial ice particle radius is \qty{2}{\mu m}, while the initial droplet size correspond to a dry aerosol particle with a diameter of \qty{0.125}{\mu m}. 

For our DNSs, we used the same numerical solver as in Refs.~\cite{sardina2015continuous,fries2021key}. The Navier-Stokes equations (\ref{eq:ns}) were solved in Fourier space using fast Fourier transform. The nonlinear terms were calculated in configuration space using the de-aliasing 2/3 rule \citep{sardina2015continuous}. Time integration used a low-storage third-order Runge-Kutta method, where the terms are treated exactly by using integration factors,
while the  non-linear terms followed an Adam-Bashforth scheme. The same Runge-Kutta  scheme was used to  integrate the equations of motion (\ref{eq:advection}) for water droplets and ice particles, and their growth equations (\ref{eq:DNS:r}).  A linear interpolation scheme was used to evaluate the air velocity and supersaturation at the particle positions, while linear extrapolation was employed to calculate the condensation rates $C_\w$ and $C_\i$ in Eqs.~(\ref{eq:condrates}).  The parameter values for the DNSs are summarised in Table~\ref{tab:A_turb}.
\begin{table}[t]
	\centering 
	\caption{\label{tab:A_turb} DNS time- and length scales for the Pi chamber~\protect\citep{chen2024pi}.}
	\begin{tabular}{@{}ll@{}} 
		\hline \hline
		 Parameter & Value \\
		\hline 
     Simulation time $t_\text{DNS}$ & \qty{600}{s} \\
     Eddy-turnover time $k/ \varepsilon$ & \qty{1.6}{s} \\
    Kolmogorov time  $\tau_\eta$  & \qty{4.4e-2}{s} \\
		 Integration time step $\Delta t$ & \qty{1e-3}{s} \\
    \hline\\[-2.5ex]
    Linear domain size $L$  & \qty{2e-1}{m} \\
      Integral length scale $L_\text{int}$ & \qty{4.2e-2}{m} \\
     Taylor microscale $\lambda$ & \qty{1.6e-2}{m} \\
		Kolmogorov length $\eta$ & \qty{7.5e-4}{m} \\
		 Spatial resolution $\Delta x$ & \qty{7.8e-4}{m}  \\
		\hline\hline
\end{tabular}
\end{table}

We also performed our own DNSs for the CTGC~\citep{chen2023mixed}, for the same parameters as in Ref.~\cite{chen2023mixed}.
They advect two scalar fields, temperature and water-vapour mixing ratio (Section \ref{sec:discussion}). The parameters for these DNS runs are given in Table~\ref{tab:control_case}. We note that we used slightly larger times steps and sligthly coarser grid than \citet{chen2023mixed}. We use our DNSs for the CTGC to determine the Lagrangian correlation time of supersaturation, an input needed for the statistical model that is discussed next.
\begin{table}
	\centering 
	\caption{DNS time- and length scales for the CTGC~\citep{chen2023mixed}.}
	\begin{tabular}{@{}ll@{}} 
		\hline\hline 
		 Parameter & Value \\
		\hline 
    Simulation time  $t_\text{DNS}$ & \qty{95}{s} \\
     Eddy turnover time $k/ \varepsilon$ & \qty{2.82}{s} \\
   Kolmogorov time  $\tau_\eta$ & \qty{1.26e-1}{s} \\
		Integration time step $\Delta t$ & \qty{2.5e-3}{s} \\
    \hline
   Linear domain size $L$ & \qty{2e-1}{m} \\
    Integral length scale  $L_\text{int}$ & \qty{1.31e-1}{m} \\
   Taylor microscale   $\lambda$ & \qty{2.13e-2}{m} \\
		 Kolmogorov length $\eta$  & \qty{1.42e-3}{m} \\
		Spatial resolution $\Delta x$ & \qty{1.5625e-3}{m}\\
		\hline\hline
	\end{tabular}
	\label{tab:control_case}
\end{table}

\section{Statistical model}
\label{sec:sm}
To understand glaciation dynamics and how it is affected by small-scale turbulence one could simulate the microscopic model described in Section \ref{sec:model} for a wide range of parameters. Here we take an alternative approach: we derive a statistical model that allows us to systematically study the parameter dependencies of the glaciation process, and provides immediate insight into possible effects of small-scale turbulence. We validate the model by showing that it yields quantitative agreement with
the DNS results of \cite{chen2023mixed}. In essence, the model is a statistical model for the supersaturation, approximating Eqs.~(\ref{eq:ns}, \ref{eq:supsat}, \ref{eq:advection}), while the Eqs.~\eqref{eq:DNS:r} for particle radii remain the same. The derivation of the model rests on two assumptions:
\begin{itemize}
    \item[A1.] The supersaturation statistics along water-droplet, ice-particle, and Lagrangian fluid paths are the same.
    \item[A2.] The supersaturation statistics are Gaussian.
\end{itemize}
Our DNSs for the CTGC and for the core of the Pi chamber show that these assumptions hold. In Section~\ref{sec:discussion} we discuss their range of validity.
To derive the model under the above assumptions, we start by decomposing the supersaturation along a particle trajectory into its mean and fluctuating parts
\begin{equation}
    s_\w = \langle s_\w \rangle + s'_\w.
\end{equation}
We use the usual notation $\langle \cdot \rangle$ for ensemble averages of physical quantities and $\cdot'$ for their fluctuating parts. 
The system is statistically homogeneous: the mean values may depend on~$t$, but not on $\bx$. Since water droplets and ice particles are Lagrangian tracers, and since the flow is incompressible, single-point Eulerian and Lagrangian statistics are the same. Therefore we can take the ensemble average of Eq.~(\ref{eq:DNS:s_w}) to obtain the evolution equation for the mean supersaturation~$\langle s_\w \rangle$:
\begin{equation} \label{eq:mean_s}
  \frac{\upd \langle s_\w \rangle}{\upd t} = - A_{2, \w} \langle C_\w \rangle - A_{2, \i} \langle C_\i\rangle + \langle f^{({s_\w})}\rangle.
\end{equation}
To close Eq.~(\ref{eq:mean_s}), we need expressions for the mean condensation rates
$\langle C_\w \rangle$ and $\langle C_\i \rangle$, which we derive in Appendix~\ref{app:cr}: 
\begin{equation}\label{eq:Cav0}
    \langle C_\w \rangle = \frac{4}{3} \pi \frac{\rho_\w}{\rho_0} n_\w \biggl< \frac{\upd r_\w^3}{\upd t} \biggr>, \qquad
    \langle C_\i \rangle = \frac{4}{3} \pi \frac{\rho_\i}{\rho_0} n_\i \biggl< \frac{\upd r_\i^3}{\upd t} \biggr>.
\end{equation}
Using Eqs.~\eqref{eq:DNS:r}, these expressions evaluate to
\begin{subequations}\label{eq:Cav0_expanded}
    \begin{align}
        \langle C_\w \rangle &=
        4 \pi \frac{\rho_\w}{\rho_0} A_{3, \w} n_\w \bigl< a_3\bigl(\tfrac{r_\w}{r_{A_3, \w}}\bigr) r_\w \bigl[(s_\w - s_{\w, \K}(r_\w) \bigr] \bigr>, \\
        \langle C_\i \rangle &=
        4 \pi \frac{\rho_\i}{\rho_0} A_{3, \i} n_\i \bigl< a_3\bigl(\tfrac{r_\i}{r_{A_3, \i}}\bigr) r_\i s_\i \bigr>.
    \end{align}
\end{subequations}
Note that averages involving particle radii also include averaging over particles, e.g.\ $\langle r_\w s_\w \rangle = \frac{1}{N_\w} \sum_{\alpha = 1}^{N_\w} \langle r_{\w, \alpha} s_{\w, \alpha} \rangle$. However for simplicity we do not introduce a special notation for such averages, except for Appendix~\ref{app:cr} where we use $\langle {\cdot} \rangle_\w$ or $\langle {\cdot} \rangle_\i$. For the Pi chamber, the average forcing $\langle f^{({s_\w})}\rangle$  in (\ref{eq:mean_s}) is given by Eq.~(\ref{eq:fs}). For the CTGC~\citep{chen2023mixed}, 
the average vanishes. 

With a model for the mean supersaturation in place, we now introduce a model for its fluctuating part. \citet{fries2024lagrangian} used the mapping closure of \citet{pope1985pdf} and ~\citet{chen1989probability} to accurately describe non-Gaussian dynamics of $s_\w'$ during the evaporation of water droplets at the cloud edge. We start from  the same model here. Since in our case the statistics of $s_\w$ is Gaussian, the mapping closure becomes a linear theory and reduces to an Ornstein--Uhlenbeck process for $s_\w'$
\begin{equation} \label{eq:SM:s}
    \upd s'_\w = - \frac{1}{\tau_{s_\w}^{({L})}} s'_\w  \, \upd t + \sqrt{\frac{2 \sigma_{s_\w}^2}{\tau_{s_\w}^{({L})}}} \, \upd W(t).
\end{equation}
Here $\upd W(t)$ are white-noise increments, while the supersaturation variance $\sigma_{s_\w}^2 = \langle s_\w'^2 \rangle$ and the correlation time $\tau_{s_\w}^{({L})}$ are the two parameters of the model. The model~\eqref{eq:SM:s} is also known as the Langevin mixing model \cite[Eq.~5.52]{pope1985pdf}. Related models have been used to describe the effect of supersaturation fluctuations on the growth of water droplets in turbulent clouds~\citep{kumala1997effect, chandrakar2016aerosol, siewert2017statistical, abade2017broadening, sardina2018broadening}. Some of them contain additional condensation terms in the equation for $s_\w'$. To understand why such terms do not matter in our case, consider a more general statistical model for the supersaturation fluctuations 
\begin{equation}\label{eq:SMC:s}
    \upd s'_\w = - A_{2,\w} \langle C_\w' \mid s_\w', t \rangle - A_{2,\i} \langle C_\i' \mid s_\w', t \rangle - \frac{s'_\w}{\tau_{s_\w}^{({L})}}   \, \upd t + \sqrt{\Ddva} \, \upd W(t).
\end{equation}
Here $\langle C'_\phi \mid s_\w', t \rangle = \langle C_\phi \mid s_\w', t\rangle - \langle C_\phi\rangle$, where  $\langle C_\phi \mid s_\w, t\rangle$  are  conditional condensation rates, and $\Ddva$ is chosen such to conserve $\sigma_{s_\w}$ [Eq.~\eqref{eq:Ddva}]. 
Appendix~\ref{app:cr} outlines how to derive Eq.~\eqref{eq:SMC:s}) using the method of~\citet{sarnitsky2022nonparametric}. 

However, the fluctuating condensation-rate contributions are negligible if the timescale of turbulent mixing $\tau_{s_\w}^{(L)}$ is much smaller than the timescales $\tau_{s_\w, \w}$ and $\tau_{s_\w, \i}$ of supersaturation evolution due to the phase change of water droplets and ice particles,
\begin{equation}
    \tau_{s_\w, \w} = \frac{\sigma_{s_\w}^2}{A_{2, \w}|\langle C_\w' s_\w' \rangle|}, \qquad
    \tau_{s_\w, \i} = \frac{\sigma_{s_\w}^2}{A_{2, \i}|\langle C_\i' s_\w' \rangle|},
\end{equation}
where $|{\cdot}|$ denotes the absolute value. Their fractions with $\tau_{s_\w}^{(L)}$ define the supersaturation Damk\"ohler numbers
\begin{equation}\label{eq:Da_s}
    \Da_{s_\w, \w} = \frac{\tau_{s_\w}^{(L)}}{\tau_{s_\w, \w}}, \qquad \Da_{s_\w, \w} = \frac{\tau_{s_\w}^{(L)}}{\tau_{s_\w, \w}}.
\end{equation}
We conclude: for small Damk\"ohler numbers, $\Da_{s_\w, \w} \ll 1$ and $\Da_{s_\w, \i} \ll 1$, one can use the model~\eqref{eq:SM:s} instead of~\eqref{eq:SMC:s}. Our model calculations confirm that the supersaturation Damk\"ohler numbers are smaller than unity for the cases studied here. 

To close the model~\eqref{eq:SM:s}, we need to provide the supersaturation variance $\sigma_{s_\w}^2$ and the correlation time $\tau_{s_\w}^{(L)}$. For the Pi chamber, the variance is given by~\citet{chen2024pi}. For the CTGC, \citet{chen2023mixed} specify the variances of temperature and the water-vapour mixing ratio, which allows us to compute $\sigma_{s_\w}^2$. The Lagrangian correlation time $\tau_{s_\w}^{({L})}$ is defined as
\begin{align}
\label{eq:tauLs}
\tau_{s_\w}^{({L})}
&= \sigma_{s_\w}^{-2} {\int_{0}^{\infty}\!\upd t\, \langle s'(t) s'(0) \rangle  },
\end{align}
where $\langle s'(t) s'(0) \rangle$ is the Lagrangian autocovariance, i.e.\ it is taken along Lagrangian trajectories. The numerical values of $\tau_{s_\w}^{(L)}$ and~$\tau_L$, are given in Appendix~\ref{app:parameters}. We note that for the cases studied here, phase change does not affect~$\tau_{s_\w}^{({L})}$ since the condensation terms are negligible in the dynamics of $s_\w'$ for small supersaturation Damk\"ohler numbers. Disregarding phase change, $\tau_{s_\w}^{(L)}$ is commonly related to the large eddy turbulent timescale $\tau_L = k/\varepsilon$:
\begin{equation}
    \tau_{s_\w}^{({L})} = \frac{C_{0, s_\w}}{C_{s_\w}} \tau_L,
\end{equation}
The Lagrangian Obukhov--Corrsin constant $C_{0, s_\w}$ comes from the Kolmogorov hypothesis extended to passive scalars and connects the correlation timescale $\tau_{s_\w}^{(L)}$ to the dissipation timescale $\sigma_{s_\w}^2 / \dr$, $\tau_{s_\w}^{(L)} = C_{0, s_\w} \sigma_{s_\w}^2 / \dr$, where $\dr = 2 \varkappa \bigl< \frac{\uppd s_\w}{\uppd x_j} \, \frac{\uppd s_\w}{\uppd x_j} \bigr>$ is the supersaturation dissipation rate. For the CTGC and Pi chamber, the values of $C_{0, s_\w}$ are $1.2$ and $1.1$, inferred from the DNS described in Section~\ref{sec:dns}.
The quantity $C_{s_\w}$ is the so called mechanical-to-scalar timescale ratio, formally defined as $C_{s_\w} = k \dr / (\sigma_{s_\w}^2 \varepsilon)$. The quantity $C_{s_\w}$ can be considered approximately constant only in specific types of flows, like the forced isotropic turbulence we deal with here~\citep{ristorcelli2006passive}. Its numerical values calculated from the DNS for the CTGC and Pi Chamber cases are $1.8$ and $2.2$ respectivly. We stress again this discussion is valid only for $\Da_{s_\w} \ll 1$ and $\Da_{s_\i} \ll 1$. In case of non-negligable supersaturation Damk\"ohler numbers, \citet{fries2024lagrangian} found that both $C_{0, s_\w}$ and $C_{s_\w}$ (denoted there as $2/C$ and $2\phi_*$) cannot be considered constant.

To numerically integrate the statistical  model, we use the Euler--Maruyama scheme with a time step of $\qty{0.05}{s}$ for the CTGC, and $\qty{0.02}{s}$ for the Pi chamber. To ensure that the numerical integration conserves the invariants~\eqref{eq:s_inv}, condensation rates are computed directly from Eqs.~\eqref{eq:Cav0} and not Eqs.~\eqref{eq:Cav0_expanded}. For the CTGC, we use $N_\w = N_\i = 10^7$ to  suppress the statistical noise for  cases 1 and 2. The choice of $N_\w$ and $N_\i$ here has no other consequences, since the model depends only on $n_\w$ and $n_\i$ in Eq.~\eqref{eq:Cav0}, which are fixed in each run. For the Pi chamber, the  number of particles is determined by the particle injection and removal process, which is implemented as described in Section \ref{sec:dns}.

Below, we refer to the deterministic limit of the statistical model, or deterministic model. It is obtained by taking the limit $\sigma_{s_\w}\to 0$ in the statistical model, which amounts to removing the white-noise term in Eq.~\eqref{eq:SM:s} and setting $s_\w' = 0$.

\begin{figure}[t]
    \begin{overpic}[width=0.8\textwidth]{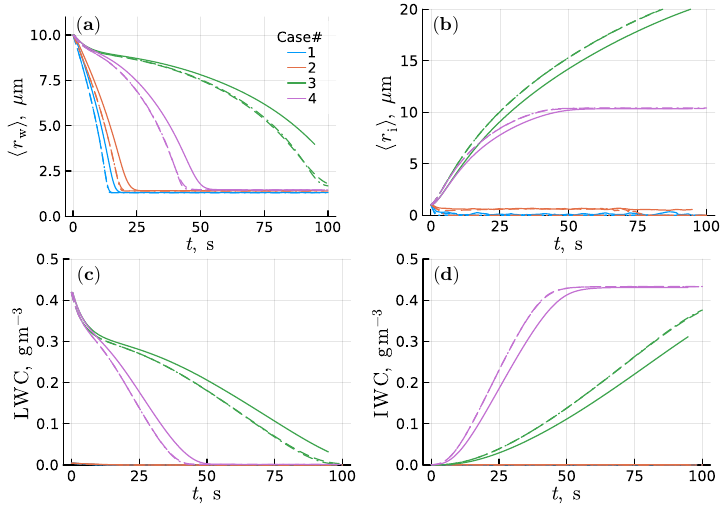}
   \end{overpic}
    \caption{\label{fig:1} Model results for ice growth in cloud-top generating cells \citep{chen2023mixed}.
    Shown are the DNS results of \citet{chen2023mixed} (solid lines) for the mean droplet radius ({\bf a}), liquid-water content LWC ({\bf b}), the mean ice-particle radius ({\bf c}) and the ice mass ({\bf d}) as functions of time. In each panel, curves for four parameter sets are shown, these parameter sets are given in Table~\ref{tab:parameters_fig1}. Only ice particles with \mbox{$r_\i > \qty{0.001}{\mu \m}$} are included in the statistics \citep{chen2023mixed}. Also shown are simulations of the statistical model (dashed lines), and of its deterministic limit (dash-dotted lines). The deterministic limit is so close to the full statistical-model results that the lines are hard to distinguish.
    }
\end{figure}

\begin{figure}[t]
       \begin{overpic}[width=0.8\textwidth]{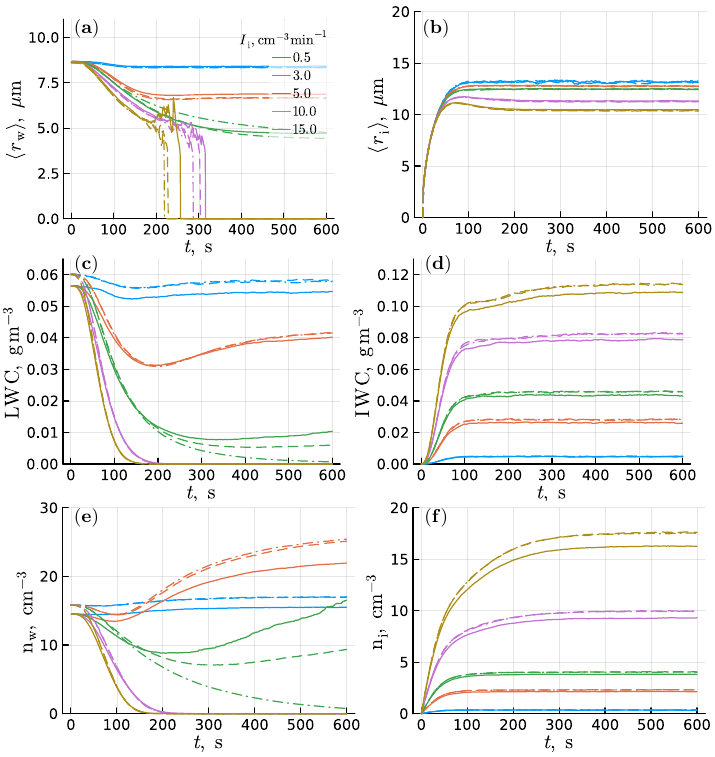}
    \end{overpic}
    \caption{\label{fig:2} Model results for ice growth in the core of the Pi chamber \citep{chen2024pi}.
    Shown are the DNS results (Section \ref{sec:dns})
    (solid lines) 
    for the mean droplet radius ({\bf a}), the liquid-water content LWC ({\bf b}), the mean ice-particle radius ({\bf c}), the ice mass ({\bf d}), water droplet concentration ({\bf e}) and ice particle concentration ({\bf f}) as functions of time. In each panel, curves for five different ice-particle injection rates [cm$^{-3}$ ${\rm min}^{-1}$] are shown, the parameter values are given in the insets. Also shown are simulations of the statistical model (dashed lines), and of its deterministic limit (dash-dotted lines). In most but not all cases, the deterministic limit is so close to the full statistical-model results that the lines are hard to distinguish.}
\end{figure}

\section{Results}
\label{sec:results}
Figure~\ref{fig:1} shows the results  for  ice growth and water evaporation due to the WBF in the CTGC \citep{chen2023mixed}.
We chose the four cases from the accompanying Replication Data~\citep{chen2023data} where the particle-size evolution is most rapid, their parameters are listed in Tables~\ref{tab:parameters_fig1} and \ref{tab:B}
in Appendix~\ref{app:parameters}.
\begin{table}[]
    \centering
    \caption{\label{tab:parameters_fig1} Parameters for the CTGC case (Fig.~\ref{fig:1}).}
    \begin{tabular}{r c c c c c c c}
        \hline\hline
        Case & $s_{\i, \text{inv}}$ & initial $\langle s_{\w} \rangle$ &  initial $r_\w$, \unit{\mu m} & initial $r_\i$, \unit{\mu m} & $n_\w$, \unit{cm^{-3}} & $n_\i$, \unit{cm^{-3}} & $\sigma_{s_\w}$ \\
        \hline
        1 & -0.080 & -0.2 & 10 & 1  & 1  & 100 & 0.017 \\
        2 & -0.023 & -0.15 & 10 & 1 & 1  & 100 & 0.017 \\
        3 & 0.42 & -0.1  & 10 & 1 & 100 & 10 & 0.016 \\
        4 & 0.42 & -0.1  & 10 & 1 & 100 & 100 & 0.016 \\
        \hline\hline
    \end{tabular}
\end{table}
Panel ({\bf a}) shows how the droplet radius shrinks because the droplets evaporate.  The radius saturates at a small value determined by the interplay between solute and curvature effects on the one hand, and evaporation on the other hand \citep{hoffmann2022}.  Shown are the DNS results of \citet{chen2023mixed} with two scalar fields, temperature and water-vapour mixing ratio (solid lines). We see that the statistical model results (dashed lines) agree very well with those of the DNS, although the droplets evaporate somewhat faster in the statistical model. Also shown are results for the deterministic limit of the model (dash-dotted lines). They are almost indistinguishable from the full statistical-model results. This shows that turbulence has no effect on the evolution of the mean droplet radius. Panel ({\bf c}) shows how the LWC decreases as the droplets evaporate, with analogous conclusions. In panel ({\bf b}), we compare how the mean ice-particle radius changes as a function of time. In agreement with the values of the invariant $s_{\i, \text{inv}}$ (Eq.~\eqref{eq:s_i_inv}, Table~\ref{tab:parameters_fig1}): the first two cases have $s_{\i, \text{inv}} < 0$ and the ice evaporates, while for the last two cases  $s_{\i, \text{inv}} > 0$ and the ice particles grow. Panel ({\bf d}) shows the IWC, which approaches a non-zero steady state when the cloud glaciates, but tends to zero in the other two cases, as expected. The fact that LWC decreases and IWC increases for cases 3 and 4 indicates that the ice particles grow at the expense of water droplets, as described by the WBF process. In summary, the main conclusion from Fig.~\ref{fig:1} is that small-scale turbulence does not affect the mean particle radius. \citet{chen2023mixed} came to the same conclusion for their base case. Comparing the deterministic limit of our statistical model to their DNS for all parameter settings 
listed in the replication data \citep{chen2023data} show that turbulence does not affect the mean particle radii for any of the cases.

Figure~\ref{fig:2} shows mean particle radii versus time for the five Pi chamber cases specified by~\citet{chen2024pi} corresponding to  different ice-particle injection rates. As specified by \citet{chen2024pi}, the mean droplet radius in Fig.~\ref{fig:1} is computed excluding droplets with radii $<3.5\,\text{$\mu$m}$. Panel ({\bf a}) reveals how the droplet radius changes. For the two highest ice-particle injection rates, the radius tends to zero. In other words, the cloud glaciates. For lower ice-particle injection rates, droplets remain in the centre of the Pi chamber at the end of the simulation. The transition to glaciation occurs for ice-injection rates between $5$ and $10$ cm$^{-3}$ ${\rm min}^{-1}$. As for the CTGC, we see that the statistical model (dashed lines) describes the DNS results (solid lines) very well, as does the deterministic limit. Also here we conclude that small-scale turbulence has little effect, at least on droplets of radii $> \qty{3.5}{\mu m}$. Panel ({\bf c}) shows how the LWC changes as a function of time. The results indicate that turbulence may change the location of the glaciation transition, as evident from the case $I_\i = \qty{5}{cm^{-3}\,min^{-1}}$ (green lines). The LWC with turbulence (the dashed line) increases at large times and the cloud remains mixed phase, whereas LWC without turbulence (the dash-dotted line) decreases and the cloud glaciates.
Approaching the glaciation transition, LWC curves show a dip, as demonstrated the clearest by the case $I_\i = \qty{3}{cm^{-3}\,min^{-1}}$ (the orange curve) near $t \sim \qty{180}{s}$. This is explained as follows. 
Initially the droplets evaporate, so the LWC falls. But as aerosol is added, the competition for water vapour increases, which leads to smaller droplets. Because these droplets do not sediment substantially, the LWC grows. Hence as the overall number of droplets keeps increasing, the
mean radius of droplets decreases and reaches a steady state, as demonstrated by panel (\textbf{e}).  Hence injecting ice particles results in a decrease of droplets size, but in an increase of droplet concentration. Panels ({\bf c}), ({\bf d}) and ({\bf f}) show how the mean ice-particle radius approaches a plateau, as does the IWC and the ice particle concentration. The differences in water contents and particle concentrations between the statistical model and the DNS are due to slight differences in the particle injection rates. The general message from Fig.~\ref{fig:2} is that small-scale turbulence has little effect on the mean radius of ice particles and droplets with $r_\w > \qty{3.5}{\mu m}$ for the parameters from~\citet{chen2024pi}, except possibly upon the timing of the glaciation transition.

\begin{figure}[t]
    \begin{overpic}[width=0.4\textwidth]{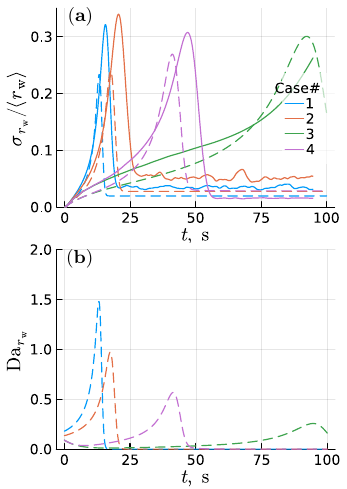}
    \end{overpic}
    \caption{\label{fig:3} (\textbf{a}) Relative dispersion of droplet radii for ice growth in cloud-top generating cells \citep{chen2023mixed}. Shown are the DNS results of \citet{chen2023mixed} (solid lines), simulations of the statistical model (dashed lines). (\textbf{b})~Statistical-model results for the Damk\"ohler number $\Da_{r_\w}$ versus time for the same cases as shown in panel ({\bf a}).}
\end{figure}

Figure~\ref{fig:3} illustrates how the fluctuations in droplet radii develop as a function of time
for the CTGC. For all four cases, the relative dispersion $\sigma_{r_\w}/\langle r_\w\rangle$ ($\sigma_{r_\w} = \sqrt{\langle r_\w '^2\rangle}$) increases rapidly, before decaying to a plateau (for the green curve the decay is not shown). 

We see that the statistical model (dashed lines) describes the DNS (solid lines) very well; the deviations are consistent with the attenuation of $\sigma_{s_\w}$ with time in the DNS of~\cite{chen2023mixed}, while in the statistical model we use $\sigma_{s_\w}$ constant in time. More importantly, in the deterministic limit $\sigma_{r_\w} = 0$ (not shown). 
We conclude: the particle-size dispersion is a consequence of small-scale turbulence, and it is well described by the statistical model. We do not show corresponding results for ice,
because
ice evaporates quickly and $\langle r_i \rangle$ reaches zero for cases~1 and~2. For cases 3 and 4, ice  grows so rapidly that the relative fluctuations in the ice-particle radii are negligible.

\begin{figure}
    \begin{overpic}[width=0.4\textwidth]{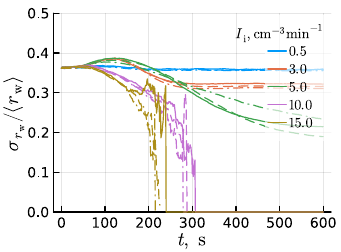}
    \end{overpic}
    \caption{\label{fig:4} Relative dispersion of droplet radii for
    the Pi chamber \citep{chen2024pi}, for
    different ice-particle injection rates [cm$^{-3}$ ${\rm min}^{-1}$]. Shown are the DNS results (Section \ref{sec:dns}, solid lines), simulations of the statistical model (dashed lines), and of its deterministic limit (dash-dotted lines). }
\end{figure}

Figure~\ref{fig:4} summarises the same for the Pi chamber, regarding the relative dispersion of droplet radii as a function of time. A major difference in this case is that particle removal and injection causes a particle-size dispersion. This is well described by the deterministic model. So here small-scale turbulence is less important for the particle-size dispersion, compared with the CTGC results in Fig.~\ref{fig:3}. Note that this observation is valid only for larger droplets with $r_\w > \qty{3.5}{\mu m}$, this cutoff also being the reason why $\sigma_{r_\w} = 0$ for glaciated clouds. Now we turn to the question of how turbulence affects smaller droplets (haze).

\begin{figure}
    \begin{overpic}[width=0.8\textwidth]{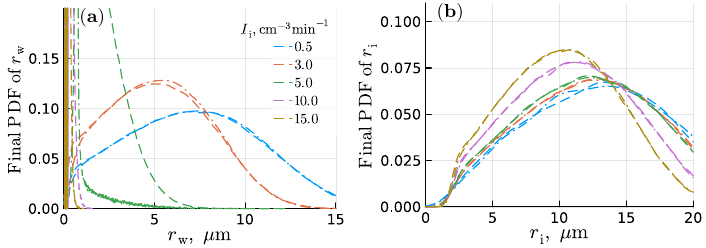}
    \end{overpic}
     \caption{\label{fig:5} 
    Final probability distributions of particle radii for the Pi chamber \citep{chen2024pi}, for different ice-particle injection rates [cm$^{-3}$ ${\rm min}^{-1}$]. Shown are  statistical-model results (solid lines), and from its
    deterministic limit (dashed). ({\bf a}) water droplets, ({\bf b}) ice particles. 
    }
\end{figure}

Figure~\ref{fig:5} shows the probability density functions (PDFs) of particle radii for the Pi chamber at large times, for water droplets [panel ({\bf a})] and for ice particles [panel ({\bf b})]. For the ice particles, statistical-model predictions and the deterministic limit agree very well, for the droplets also, but not near or after the glaciation transition. The smaller the mean particle radius the greater the shift of the deterministic limit PDF to smaller radii compared to the statistical-model PDF. Thus, we see that turbulence widens the distribution of droplet radii for smaller droplets.

\section{Discussion}\label{sec:discussion}
Figures~\ref{fig:1} and \ref{fig:2} show that the statistical-model predictions 
agree very well with the DNS results. We now explain why this is the case here, and under which circumstances the model may fail. To this end we study the limits of validity of the statistical model, which we derived from the assumptions A1 and A2 listed in~Section~\ref{sec:sm}. 

The first assumption, A1,  is that the supersaturation statistics along water-droplet, ice-particle, and Lagrangian fluid paths are the same. We expect this to hold when the Damk\"ohler numbers $\Da_{s_\w, \w}$ and $\Da_{s_\w, \i}$ [Eq.~\eqref{eq:Da_s}] are small, and when water droplets and ice particles are initially well-mixed. Under these conditions, turbulence transports water vapour to and from particles faster than phase change occurs. Hence the particles have no time to form a supersaturation field in their vicinity, which might differ from  particle-free regions of the flow; all the particles experience the same supersaturation statistics.

The second assumption, A2, is that the supersaturation statistics are Gaussian. In a homogeneous system (no scalar gradients~\citep{Pumir+91, warhaft1991probability, Gollub+91}), the steady-state distribution of a passive scalar in isotropic homogeneous turbulence is Gaussian \citep{jayesh1992probability, warhaft2000passive, kumar2012extreme, kumar2014lagrangian}. Non-Gaussian tails that may be present in transient mixing~\citep{Villermaux:2019} disappear as the steady state is approached. 

In summary, the statistical model from Section~\ref{sec:sm} can be justified when
the Damk\"ohler numbers $\Da_{s_\w, \w}$ and $\Da_{s_\w, \i}$ are small. The cases summarised in Section \ref{sec:results} all have Damk\"ohler numbers smaller than unity, explaining why the statistical model works so well. On the other hand, it is worth noting that the interaction between turbulence and phase change can be more intricate and harder to describe at larger Damk\"ohler numbers. This is in line with the findings of~\citet{fries2024lagrangian}, and requires more refined mapping-closure approximations \citep{chen1989probability,pope1985pdf}.

Next, we quantify the role of turbulence on the dynamics of mean supersaturation and upon the mean particle radii. In the framework of the statistical model, the question is  
how $\sigma_{s_\w}$ -- the measure of turbulence intensity -- enters the dynamical equations for $\langle s_\w \rangle$ and $\langle r_\w \rangle$. To this end we need to introduce two 
more Damk\"ohler numbers [besides the supersaturation Damk\"ohler numbers (\ref{eq:Da_s})]:
\begin{equation}\label{eq:Da_r_phi}
    \Da_{r_\w} = \frac{\tau_{s_\w}^{(L)}}{\tau_{r_\w}}, \qquad
    \Da_{r_\i} = \frac{\tau_{s_\w}^{(L)}}{\tau_{r_\i}}.
\end{equation}
They are associated 
with the timescales $\tau_{r_\w}$ and $\tau_{r_\i}$ of the particle-size evolution:
\begin{equation}\label{eq:tau_r_phi}
    \tau_{r_\w} =
    \frac{\langle r_\w\rangle^2}{\bigl| \tfrac{{\rm d}}{{\rm d}t}\langle r_\w^2 \rangle \bigr|}, \qquad
    \tau_{r_\w} =
    \frac{\langle r_\i\rangle^2}{\bigl| \tfrac{{\rm d}}{{\rm d}t}\langle r_\i^2 \rangle \bigr|}
\end{equation}
We note that these Damk\"ohler numbers scale as $\sim 2 A_{3} s_\phi \tau_{s_\w}^{(L)}/\langle r_\phi\rangle^2$ ($\phi = \w$ or $\phi = \i$). Therefore they are larger
for smaller particles.

Figures~\ref{fig:1} and \ref{fig:2} imply that small-scale turbulence has only a weak
effect on the evolution of the average supersaturation. To explain this, we look at the evolution equation~\eqref{eq:mean_s} for $\langle s_\w \rangle$. It is written in terms of the mean condensation rates $\langle C_\w \rangle$ and $\langle C_\i \rangle$. For the CTGC (no particle injection or removal), we can estimate $\langle C_\w \rangle$ and $\langle C_\i \rangle$~as
\begin{equation}\label{eq:cwav_da}
    \begin{aligned}
        \langle C_\phi \rangle
        &= 4 \pi \frac{\rho_\phi}{\rho_0} A_{3, \phi} n_\phi \bigl(\langle r_\phi \rangle \langle s_\phi \rangle + \langle r'_\phi s'_\phi \rangle\bigr) 
        \approx 4 \pi \frac{\rho_\phi}{\rho_0} A_{3, \phi} n_\phi \langle r_\phi \rangle \langle s_\phi \rangle \biggl(1 + \frac{1}{2} \frac{\sigma_{s_\phi}^2}{\langle s_\phi \rangle |\langle s_\phi \rangle|} \Da_{r_\phi} \biggr),
    \end{aligned}
\end{equation}
where $\phi = \w$ or $\phi = \i$, as before. 
These approximations are derived in Appendix~\ref{app:moments} for  a simplified particle-growth model, and assuming small particle Damk\"ohler numbers, small supersaturation Damk\"ohler numbers, and narrow particle-size distributions.
Equation~(\ref{eq:cwav_da}) shows that the turbulence term $\langle r_\phi' s_\phi' \rangle$ is proportional to the particle Damk\"ohler numbers. It can therefore be neglected when $\Da_{r_\w}$ and $\Da_{r_\i}$ are small, when the particles are large enough. Neglecting $\langle r_\phi' s_\phi' \rangle$ corresponds to the decoupling between the fluctuations of radii and supersaturation when $s_\phi'$ evolves much faster than $r_\phi'$ for large particles, so that $r'_\phi$ can respond only to the slow evolving $\langle s_\phi \rangle$, but not $s_\phi'$.  We note that Eq.~(\ref{eq:cwav_da}) fails near $\langle s_\phi \rangle \approx 0$, when the turbulence term $\langle r_\phi' s_\phi' \rangle$ becomes comparable to $\langle r_\phi \rangle \langle s_\phi \rangle$. In practice this affects most the stationary state where the system spends significant time at $s_\i \approx 0$, i.e. after  glaciation has  happened. 

Now consider the evolution of the average particle radii. In Appendix~\ref{app:moments} we derive an approximate equation for $\langle r_\phi \rangle$ under the same assumptions used to derive Eq.~\eqref{eq:cwav_da}
\begin{equation}
    \label{eq:mre}
    \frac{\upd \langle r_\phi \rangle}{\upd t}
    \approx A_{3, \phi} \frac{\langle s_\phi \rangle}{\langle r_\phi \rangle} \biggl(1 - \frac{1}{2} \frac{\sigma_{s_\phi}^2}{\langle s_\phi \rangle | \langle s_\phi \rangle |} \Da_{r_\phi} \biggr).
\end{equation}
This result shows that small-scale turbulent fluctuations are negligible for small particle Damk\"ohler numbers. The physical reason why the mean radius dynamics is unaffected by turbulent fluctuations of supersaturation is the same decoupling between the dynamics of $r_\phi'$ and $s_\phi'$ discussed in the previous paragraph. Equation~(\ref{eq:mre}) comes with the same caveat as above, it fails near $s_\phi \approx 0$. We stress that the statistical model correctly describes the mean radius even when $s_\phi \approx 0$.

In conclusion, Figs.~\ref{fig:1} and \ref{fig:2}, as well as Eqs.~(\ref{eq:cwav_da}) and (\ref{eq:mre}) show that turbulence has at best a weak effect upon the evolution of the mean particle size at small Damk\"ohler numbers. This implies that small-scale turbulence neither affects the
WBF process nor the resulting glaciation time, defined as the time it takes for the ice-particle mass fraction IWC/(IWC+LWC+$q_\v\rho_0$) to reach 90\% \citep{chen2024pi}.

On the other hand, Figure~\ref{fig:3}(\textbf{a}) shows that small-scale turbulence has an effect upon the fluctuations  $\sigma_{r_\phi}^2 = \langle r_\phi'^2 \rangle$ of the particle radii, as \citet{chen2023mixed} concluded in their DNS study. Our model can explain this as follows. Equations~\eqref{eq:DNS:r},  $\upd r_\phi / \upd t \sim A_{3, \phi} s_\phi / r_\phi$ imply that particles evaporate more rapidly the smaller they are. As a consequence, any small differences in radii of evaporating particles amplify in a deterministic fashion, explaining the rapid growth of the variance in Fig.~\ref{fig:3}. However, when the initial variance vanishes as for the CTGC,  turbulent supersaturation fluctuations are required to initially widen the particle-size distribution, triggering the amplification.
The widening of $\sigma_{r_\phi}$ happens for water droplets, but not for ice particles. Since the ice particles  grow, the variance $\langle r_\i'^2 \rangle$ shrinks instead. We note, finally, that the peaks of $\sigma_{r_\w} / \langle r_\w \rangle$ in Fig.~\ref{fig:3} coincide with the peaks of the droplet Damk\"ohler number $\Da_{r_\w}$ [Fig.~\ref{fig:3}(\textbf{b})], confirming that larger particle Damk\"ohler numbers imply  stronger coupling between $s_\phi'$ and~$r_\phi'$.

The above reasonings regarding the effect of small-scale turbulence on the average particle radii and their fluctuations have to be adjusted when discussing the core of the Pi chamber (Fig.~\ref{fig:4}), because we did not account for the particle injection and removal. In this case the particle-size distribution of larger droplets ($r_\w > \qty{3.5}{\mu m}$ as in Fig.~\ref{fig:4}) is dominated not by turbulence, but by injection of small particles and removal of larger ones, so small-scale turbulence has a weaker effect, compared with the CTGC. However, if we look at the PDF of $r_\w$ for all droplet sizes [Fig.~\ref{fig:5}(\textbf{a})], we see that turbulence has no influence only on cases with low ice injection rates. These cases correspond to larger particle sizes and smaller droplet Damk\"ohler number $\Da_{r_\w}$. Once the ice injection rate increases and the cloud glaciates, only small or even unactivated droplets remain. Their $\Da_{r_\w}$ is larger and they are sensitive to supersaturation fluctuations, which widen droplet size distribution compared to the case without turbulence. The distributions of ice particle size [Fig.~\ref{fig:5}(\textbf{b})] is unaffected by turbulence, because ice remains large and $\Da_{r_\i}$ is small.

At larger Damk\"ohler numbers, small-scale turbulence could have a larger effect, in particular for spatially inhomogeneous initial conditions \citep{fries2024lagrangian}. Consider for example increasing the simulation-box size in the DNS to increase the size of the unresolved turbulent scales. This increases the Damk\"ohler numbers (\ref{eq:Da_r_phi}), causing small-scale turbulence to be more important. 
In~\citep{chen2023mixed,chen2024pi} the linear size of the simulation domain was about $\qty{0.2}{m}$, the Damk\"ohler number Da$_{r_\w}$ does not exceed 1.5 [Fig.~\ref{fig:3}({\bf b})].
In order to get Damk\"ohler numbers of order 15, where turbulence is expected to have a stronger effect \citep{fries2024lagrangian}, the linear size should be at least about $\qty{20}{m}$ (since $\tau_L \sim (L^2/\varepsilon)^{1/3}$).  But as discussed above, the statistical model may fail when the Damk\"ohler numbers become too large. In this case more refined approximations for the condensation rates are needed \citep{fries2024lagrangian}.
It could also be of interest to formulate models aimed to describe the large-Da limit, as used in dense evaporating sprays \citep{villermaux2017fine}.

Here we considered high number densities of ice particles, typical for deeper clouds. 
Polar stratus clouds tend to have lower ice-particle number densities, of the order $10^{-3}\,$cm$^{-3}$ \citep{morrison2012}. In this case it is hard to reach the steady state with DNS (although it is possible for the statistical model). Figure~7 of \citet{chen2023mixed} shows results for the initial growth of ice particles at lower ice-particle number density 
(the lowest value is $8 \times 10^{-3}\,$cm$^{-3}$).  We performed our own DNSs for these cases, and find good agreement with the statistical-model results, and with the deterministic limit of the model (not shown). 

\citet{abade2024persistent} found
that turbulence increases cloud glaciation times, and that ice particles and supercooled droplets experience different supersaturation fluctuations, in apparent contradiction with the DNS of \cite{chen2023mixed}, and with our statistical-model results. This discrepancy is explained by the approximations for condensation and deposition rates used in \cite{abade2024persistent}, where the influence of neighbouring particles (ice particles or water droplets) is not taken into account, hindering the WBF process. 

\citet{korolev2022how} speculated that ice particles and
water droplets may be locally unmixed at small scales. In this
case one expects the supersaturation distributions to be
non-Gaussian \citep{fries2024lagrangian}, causing the statistical model to fail, in the form used here. Instead, one must rely on improved approximations of the supersaturation dynamics, such as mapping-closure approximations \citep{chen1989probability,pope1985pdf,fries2024lagrangian}.
In this case we certainly expect the deterministic model to be inaccurate, in particular for the glaciation time. The question is by how much local unmixing delays  glaciation, and how much turbulence contributes to the acceleration of it. This question also relates to a recent field campaign in which supercooled clouds were seeded with ice nuclei to initiate the WBF \citep{henneberger2023seeding}. For these experiments, turbulence appears to be crucial to expose the newly nucleated ice crystals to water droplets \citep{omanovic2024evaluating}.

\section{Conclusions}
\label{sec:conclusions}
We analysed the effect of small-scale turbulence on the glaciation process in mixed-phase clouds using a statistical model similar to those used to describe droplet evaporation
in warm clouds. We found that the model describes DNS results for the glaciation process very well, for the parameters
in~\citep{chen2023mixed} corresponding to a cloud-top generating cell and also for the parameters specified in~\citep{chen2024pi} corresponding
to the core of the Michigan Pi cloud chamber. 

The statistical-model analysis shows that small-scale turbulence has an overall small effect on ice growth. Small-scale turbulence affects the evaporation of water droplets in the CTGC in that it initialises the growth of the droplet-size variance. In the core of the Pi chamber, for the parameters specified by \citet{chen2024pi}, small-scale turbulence matters only close to the glaciation transition, where the water droplets are very small. When the corresponding time scale for droplet evaporation decreases
so that it is of the same order of magnitude as the Lagrangian mixing time, then small-scale turbulence may have a stronger effect.

Our calculations show more generally that the effect of  turbulence on glaciation is expected to be larger at larger Damk\"ohler numbers, i.e., on larger spatial scales $L$. 
Since mixed-phase clouds exhibit a large range of vertical and horizontal length scales, covering shallow stratiform clouds of just hundred meters depth and a horizontal extent of several tens to hundreds of kilometers to deep convective clouds extending across the entire troposphere, the potential for small-scale turbulence to affect the glaciation process in these clouds needs to be acknowledged. This is especially true since most models used in the atmospheric sciences (from large-scale global circulation models to comparably high-resolution large-eddy simulation models) do not represent the effects of small-scale turbulence on cloud microphysical processes such as the WBF process on length scales smaller than at least few tens to a couple of hundred meters \text{(i.e., their grid spacing)}.

A statistical model, such as the one presented here, may not only enable us to assess the effects of small-scale turbulence on the WBF process on larger length scales in a follow-up study, but could also be the basis of a parameterisation to represent the effects of small-scale turbulence in larger-scale models, which still struggle to represent mixed-phase clouds, and especially the coupling of cloud microphysics and turbulence \citep{vignon2021}.

{\em Acknowledgments.}
BM, FH, and AP thank E. Bodenschatz for numerous discussions about the role of turbulence in atmospheric clouds.
B. Mehlig and G. Sartnitsky were supported by Vetenskapsr\aa{}det (grant no.~2021-4452). F. Hoffmann was supported by the Emmy Noether program of the German Research Foundation (DFG) under Grant HO~6588/1-1.
Gaetano Sardina was supported by Vetenskapsr\aa{}det (grant no.~2023-2026) and  by ERC grant MixClouds 101126050 funded by the European Union. Views and opinions expressed are however those of the author(s) only and do not necessarily reflect those of the European Union or the European Research Council Executive Agency. Neither the European Union nor the granting authority can be held responsible for them. The computations  were enabled by resources provided by the National Academic Infrastructure for Supercomputing in Sweden (NAISS), partially funded by the Swedish Research Council through grant agreement no. 2022-06725.


%
\vfill\eject

\appendix
\section{Model parameters}
\label{app:parameters}
In this Appendix we summarise the details necessary to understand the definition and values 
of the model parameters used in the main text, for the CTGC~\citep{chen2023mixed} and for the Pi chamber~\citep{chen2024pi}.

{\em Saturation pressures.}
The saturation pressures $p_{\v,\w}$ and $p_{\v,\i}$ are functions of only temperature. For the Pi chamber we take~\citep{WMO_2021}
\begin{subequations}\label{eq:pvs}
    \begin{align}
      p_{\v, \w}(T) \label{eq:pw}
      &= \qty{611.2}{Pa} \ \exp\biggl(17.62 \, \frac{T - \qty{273.15}{K}}{T - \qty{30.03}{K}} \biggr), \\[1ex]
      p_{\v, \i}(T) \label{eq:pi}
      &= \qty{611.2}{Pa} \ \exp\biggl(22.46 \, \frac{T - \qty{273.15}{K}}{T - \qty{0.53}{K}} \biggr).
    \end{align}
\end{subequations}
For CTGC we take 
(Chen, private communication)
\begin{align}
    p_{\v,\w}(T) &= \qty{2.53e11}{Pa} \, \exp\biggl(-\frac{\qty{5.42e3}{K}}{T} \biggr), \\
    p_{\v,\i}(T) &= \qty{3.41e12}{Pa} \, \exp\biggl(-\frac{\qty{6.13e3}{K}}{T} \biggr).
\end{align}

{\em Parameters $A_{3,\phi}$ and $r_{A_{3,\phi}}$.}
To derive expressions~\eqref{eq:A_3} and \eqref{eq:r_A_3}, we start from the well-known form of the particle growth equations~\citep{Mordy_1959, yau1989short}:
\begin{subequations}\label{eq:r_full}
\begin{align}
  \frac{\upd r_\w^2}{\upd t}
  &= 2 \hat A_{3, \w}(r_\w) \, \bigl[ s_\w - s_{\w, \K}(r_\w) \bigr], \\
  \frac{\upd r_\i^2}{\upd t}
  &= 2 \hat A_{3, \i}(r_\i) \, a_3(r_\i) \, s_\i.
   \end{align}
\end{subequations}
To derive expressions~\eqref{eq:A_3} and \eqref{eq:r_A_3} we need to show that $\hat A_{3, \phi}(r_\phi) = A_{3, \phi} a_3(r_\phi / r_{A_3, \phi})$. The radius-dependent function $\hat A_{3, \phi}$ in Eq.~\eqref{eq:r_full} is given by
\begin{subequations}\label{eq:full_A_3}
    \begin{align}
      \hat A_{3, \phi} \label{eq:A_3_full}
      &= \biggl[ \biggl( \frac{L_\phi(T_0)}{R_\v T_0} - 1 \biggr) \frac{R_\a}{c_p} \frac{\rho_\phi L_\phi(T_0)}{\varkappa'_{T, \phi} p_0} + \frac{\rho_\phi R_\v T_0}{\varkappa'_{q_\v, \phi} p_{\v, \phi}(T_0)}\biggr]^{-1}, \\
      \varkappa'_{T, \phi}(r_\phi) \label{eq:kappa_T_prime}
      & = \varkappa_T \biggl( \frac{r_\phi}{r_\phi + \Delta_T} + \frac{\varkappa_T}{r_\phi \alpha_{T, \phi}} \sqrt{\frac{2\pi}{R_\a T_0}} \biggr)^{\!-1}, \\
      \varkappa'_{q_\v, \phi}(r_\phi) \label{eq:kappa_q_v_prime}
      & = \varkappa_{q_\v} \biggl( \frac{r_\phi}{r_\phi + \Delta_{q_\v}} + \frac{\varkappa_{q_\v}}{r_\phi \alpha_{q_\v, \phi}} \sqrt{\frac{2\pi}{R_\v T_0}} \biggr)^{\!-1}.
    \end{align}
\end{subequations}
This form for $\hat A_{3, \phi}$ was used by \cite{chen2023mixed} and \cite{chen2024pi}, however we note that~\cite{Lamb:2011} provide a different formula that involves the K\"ohler correction function $s_{\w, \K}$.
Here $\Delta_T$ and $\Delta_{q_\v}$ 
are scales of the order of the mean free path in air, $\lambda_\a$, parameterising kinetic corrections. The parameters $\alpha_{T, \phi}$ and $\alpha_{q_\v, \phi}$ are thermal and condensation accommodation coefficients. We take their values from~Refs.~\citep{pruppacher1997microphysics, Andreas_2005, Soelch_2010},
\begin{equation}
  \begin{gathered}
    \Delta_T = \qty{2.16e-7}{m}, \quad \Delta_{q_\v} = \qty{0.87e-7}{m}, \\
    \alpha_{T, \w} = \alpha_{T, \i} = 0.7,
    \quad \alpha_{q_\v, \w} = 0.036, \quad \alpha_{q_\v, \i} = 0.5.
  \end{gathered}
\end{equation}
For these values, the dominant radius-dependent term in Eq.~\eqref{eq:A_3_full} is the one with $\varkappa_{q_\v, \phi}'$. Thus we the ignore kinetic corrections and the temperature accommodation correction, effectively setting $\Delta_T = \Delta_{q_\v} = 0$ and $\varkappa_{T, \phi} = \varkappa_T$. In addition, we observe that $L_\phi / (R_\v T_0) \gg 1$ and proceed to derive $\hat A_{3, \phi} = A_{3, \phi} a_3(r_\phi / r_{A_{3, \phi}})$:
\begin{equation}
    \begin{aligned}
      \hat A_{3, \phi}
      &= \biggl[ \biggl( \frac{L_\phi(T_0)}{R_\v T_0} - 1 \biggr) \frac{R_\a}{c_p} \frac{\rho_\phi L_\phi(T_0)}{\varkappa_{T, \phi} \, p_0} + \frac{\rho_\phi R_\v T_0}{\varkappa_{q_\v, \phi}\, p_{\v, \phi}(T_0)}\biggr]^{-1} \\
      &= \biggl( \frac{R_\a}{R_\v} \frac{\rho_\phi L_\phi^2(T_0)}{\varkappa_T c_p T_0 p_0} + \frac{\rho_\phi R_\v T_0}{\varkappa_{q_\v} p_{\v, \phi}(T_0)} + \frac{\rho_\phi \sqrt{2\pi R_\v T_0}}{r_\phi \alpha_{q_\v, \phi} p_{\v, \phi}(T_0)} \biggr)^{-1} \\
      &= \biggl( \frac{1}{A_{3, \phi}} + \frac{1}{A_{3, \phi}} \frac{r_{A_{3, \phi}}}{r_\phi}\biggr)^{-1} = A_3 \, a_3(r / r_{A_{3, \phi}}).
    \end{aligned}
\end{equation}
The last row yields Eqs.~\eqref{eq:A_3} and \eqref{eq:r_A_3}.

{\em Supersaturation variance and correlation time.}
Now we describe how we obtained~$\sigma_{s_\w}$ and $\tau_{s_\w}^{(L)}$ for the CTGC~\citep{chen2023mixed} and for the Pi chamber~\citep{chen2024pi}.

The value of~$\sigma_{s_\w}$ is specified for the Pi chamber, but for the CTGC only the standard deviations $\sigma_{q_\v}$ and $\sigma_{T}$ of the mixing ratio and temperature are provided. We reconstruct the value of $\sigma_{s_\w}$ from the data provided in the supplementary material for~\cite{chen2023mixed}. Recall that $A_4(T) = p_{\v, \w}(T) / p_{\v, \i}(T)$, so expanding $A_4$ near $T = \langle T \rangle$ we express $s_\i$ as
\begin{equation}
    s_\i
    = A_4(T) (s_\w + 1) - 1
    = \biggl[A_4(\langle T \rangle) + T' \frac{\upd A_4}{\upd T}\bigg|_{T = \langle T \rangle} + \ldots \biggr] (s_\w + 1) - 1.
\end{equation}
After averaging this expression we can express the correlation $\langle s'_\w T' \rangle$, Eq.~\eqref{eq:swT}. At the same time, we can compute the correlations $\langle s_\w' T' \rangle$ and 
$\langle s_\w'^2\rangle$ straight from Eq.~\eqref{eq:linear_s_w}, taking $T_0 = \langle T \rangle$, $p_0 = \langle p \rangle$, $s_{\w, 0} = \langle s_\w \rangle$. As before, we note that in our case the pressure term is negligible within the Ober\-beck–Bous\-sinesq approximation. Overall we get a system of three equations:
\begin{subequations}
    \begin{align}
        \langle s_\w' T' \rangle \label{eq:swT}
        &= \Bigl[ \bigl(\langle s_\i \rangle + 1\bigr) - A_4(\langle T \rangle) \bigl(\langle s_\w \rangle + 1\bigr) \Bigr] \Bigm/ \frac{\upd A_4}{\upd T}\Big|_{T = \langle T \rangle}, \\
        \langle s_\w' T' \rangle
        &= \frac{R_\v}{R_\a} \frac{\langle p \rangle}{p_{\v,\w}(\langle T \rangle)} \langle q'_\v T' \rangle - (1 + \langle s_\w \rangle) \frac{L_\w(\langle T \rangle)}{R_\v \langle T \rangle^2} \langle T'^2 \rangle, \\
        \langle s_\w'^2 \rangle
        &= \biggl[ \frac{R_\v}{R_\a} \frac{\langle p \rangle}{p_{\v,\w}(\langle T\rangle )} \biggr]^2 \langle q_\v'^2 \rangle + \biggl[(1 + \langle s_\w \rangle) \frac{L_\w(\langle T \rangle)}{R_\v \langle T \rangle^2} \biggr]^2 \langle T'^2 \rangle \notag \\
        &\quad - 2 (1 + \langle s_\w \rangle ) \frac{R_\v}{R_\a} \frac{\langle p \rangle}{p_{\v,\w}(\langle T \rangle)} \frac{L_\w(\langle T \rangle)}{R_\v \langle T \rangle^2} \langle q_\v' T' \rangle.
    \end{align}
\end{subequations}
We solve this system to express $\sigma_{s_\w}^2 = \langle s_\w'^2 \rangle$ as a function of $\sigma_{q_\v}^2 = \langle q_\v'^2 \rangle$, $\sigma_T^2 = \langle T'^2 \rangle$, $\langle s_\w \rangle$, $\langle s_\i \rangle$, $\langle T \rangle$, $\langle p \rangle$; these datasets are provided in~\cite{chen2023data}.

The calculation of $\tau_{s_\w}$ as the parameter of the model~\eqref{eq:SM:s} follows the procedure described in \citep[Eq.~5.2 and Appendix F]{sarnitsky2021inference}.

{\em Thermodynamic parameter values.}
The thermodynamic parameters for the CTGC case are given in  Table~\ref{tab:A}. They were taken from either directly from \citep{chen2023mixed} and its Replication Data~\cite{chen2023data}, or from the previous papers in the series~\cite{grabowski2011droplet, chen2018bridging, chen2020impact}. For the reference value of supersaturation $s_{\w, 0}$ we take the mean of $s_\w = 0$ (water vapour saturated w.r.t.\ water) and $s_\w = 1 / A_4 - 1 $ (water vapour saturated w.r.t.\ ice):
\begin{equation}
    s_{\w, 0} = \frac{1}{2}\Big( \frac{1}{A_4} - 1 \Big).
\end{equation}
This value of $s_{\w, 0}$ should be the a good general choice for describing WBF process in which supersaturation lies between these two points.

The parameter values for the Pi chamber \citep{chen2024pi} are summarised in Table~\ref{tab:B}.
  For the diffusivities of temperature and water-vapour mixing ratio, $\varkappa_T$ and~$\varkappa_{q_\v}$, as well as the kinematic visocity $\nu$ of air, we used~\citep{Massman_1998,Massman_1999}:
  \begin{subequations}
  \label{eq:vkk} 
\begin{align}
  \nu(p, T)
  &= \qty{1.327e-5}{\frac{m^2}{s}}\, \frac{\qty{101325}{Pa}}{p} \, \biggl(\frac{T}{\qty{273.15}{K}}\biggr)^{\!1.81}, \\[1ex]
  \varkappa_T(p, T)
  &= \qty{1.869e-5}{\frac{m^2}{s}}\, \frac{\qty{101325}{Pa}}{p} \, \biggl(\frac{T}{\qty{273.15}{K}}\biggr)^{\!1.81}, \\[1ex]
  \varkappa_{q_\v}(p, T)
  &= \qty{2.178e-5}{\frac{m^2}{s}}\, \frac{\qty{101325}{Pa}}{p} \, \biggl(\frac{T}{\qty{273.15}{K}}\biggr)^{\!1.81}.
\end{align}
\end{subequations}
The value for $s_{\w, 0}$ was calculated from~\eqref{eq:s_w_0} from the values of $q_{\v, 0}$ and $T_0$ specified by~\cite{Wang:2022}.

\begin{table}[p]
  \centering 
  \caption{\label{tab:A} Model parameters for Pi chamber~\protect\citep{chen2024pi}.}
  \begin{tabular}{@{}cl@{}}
    \hline\hline
    Parameter & Value \\
    \hline
    $R_\a$ & \qty{287.05}{J/(kg\,K)} \\
    $R_\v$ & \qty{461.52}{J/(kg\,K)}, \\
    $c_\p$ & \qty{1005}{J/(kg\,K)} \\
    \hline
    $T_{0}$ & \qty{265.63}{K} \\
    $p_{0}$ & \qty{1e5}{Pa} \\
    $\rho_0$ & \qty{1.311}{kg/m^3}\\
    $V$ & \qty{8e-3}{m^3} \\
    $s_{\w, 0}$ & \qty{5.253e-2}{} \\
    $q_{\v, 0}$ & \qty{2.28e-3}{kg/kg} \\
    $\nu$ & \qty{1.278e-5}{m^2 / s} \\
    $\varkappa_T$ &\qty{1.800e-5}{m^2 / s} \\
    $\varkappa_{q_\v}$ &\qty{2.098e-5}{m^2 / s} \\
    $\varkappa$ & \qty{1.944e-5}{m^2 / s}  \\
    $\rho_0$ & \qty{1.311}{kg / m^3} \\
    $\rho_\w$ & \qty{1000}{kg / m^3} \\
    $\rho_\i$ & \qty{917}{kg / m^3} \\
    \hline
    $\alpha_{q_\v, \w}$ & \qty{0.036}{} \\
    $\alpha_{q_\v, \i}$ & \qty{0.5}{} \\
    $r_{A_{3, \w}}$ & \qty{2.805e-6}{m} \\
    $r_{A_{3, \i}}$ & \qty{0.1906e-6}{m} \\
    $r_\d$ & \qty{0.0625e-6}{m} \\
    $r_{\w, \text{initial}}$ & \qty{0.0625e-6}{m} \\
    $r_{\i, \text{initial}}$ & \qty{2e-6}{m} \\
    $\kappa$ & 1.12  \\
    \hline
    $A_{2, \w}$ & \qty{664.8}{} \\
    $A_{2, \i}$ & \qty{691.0}{} \\
    $A_{3, \w}$ & \qty{40.08e-12}{m^2 / s} \\
    $A_{3, \i}$ & \qty{38.27e-12}{m^2 / s} \\
    $A_4$ & \qty{1.078}{} \\\hline
    $k_{\infty, \w}$ & \qty{1.233e8}{m^{-1}s^{-1}} \\
    $k_{\infty, \i}$ & \qty{1.131e8}{m^{-1}s^{-1}} \\ 
    $H$ & \qty{0.2}{m} \\
    $I_\w$ & $10 \times \frac{10^6}{60} \, \unit{\m^{-3} s^{-1}}$\\
    
    \hline
    $\sigma_T$ & \qty{0.72}{K} \\
    $\sigma_{q_\v}$ & \qty{0.165e-3}{} \\
    $\sigma_{s_\w}$ & \qty{2.047e-2}{} \\
    $\tau_{s_\w}^{(L)}$ & \qty{0.755}{s} \\
    $\tau_{s_\w,{\rm force}}$ & \qty{60}{s} \\
    $s_{\w,{\rm force}}$ & \qty{5.253e-2}{} \\
    \hline\hline
  \end{tabular}
\end{table}

\begin{table}[p]
  \centering 
  \caption{  \label{tab:B} Model parameters for CTGC \protect\citep{chen2023mixed}.}
  \begin{tabular}{@{}cl@{}} 
    \hline\hline
    Parameter & Value \\
    \hline
    $R_\a$ & \qty{287}{J/(kg\,K)} \\
    $R_\v$ & \qty{467}{J/(kg\,K)}, \\
    $c_\p$ & \qty{1005}{J/(kg\,K)} \\
    \hline
    $T_{0}$ & \qty{259.53}{K} \\
    $p_{0}$ & \qty{57160}{Pa} \\
    $\rho_0$ & \qty{0.7674}{kg / m^3} \\
    $V$ & \qty{8e-3}{m^3} \\
    $s_{\w, 0}$ & \qty{-6.298e-2}{} \\
    $q_{\v, 0}$ & \qty{2.171e-3}{kg/kg} \\
    $\nu$ & \qty{1.6e-5}{m^2 / s} \\
    $\varkappa_T$ & \qty{2.22e-5}{m^2 / s} \\
    $\varkappa_{q_\v}$ & \qty{2.55e-5}{m^2 / s} \\
    $\varkappa$ & \qty{2.379e-5}{m^2 / s} \\
    $\rho_\w$ & \qty{1000}{kg / m^3} \\
    $\rho_\i$ & \qty{917}{kg / m^3} \\
    \hline 
    $\alpha_{q_\v, \w}$ & \qty{0.036}{} \\
    $\alpha_{q_\v, \i}$ & \qty{0.036}{} \\
    $r_{A_{3, \w}}$ & \qty{3.313e-6}{m} \\
    $r_{A_{3, \i}}$ & \qty{3.182e-6}{m} \\
    $r_\d$ & \qty{1e-6}{m} \\
    $\kappa$ & 0.3 \\
    \hline
    $A_{2, \w}$ & \qty{621.6}{} \\
    $A_{2, \i}$ & \qty{646.5}{} \\
    $A_{3, \w}$ & \qty{2.945e-11}{m^2 / s} \\
    $A_{3, \i}$ & \qty{2.696e-11}{m^2 / s} \\
    $A_4$ & \qty{1.144}{} \\
    \hline
    $\sigma_T$ & \qty{0.143}{K} \\
    $\sigma_{q_\v}$ & \qty{4.5e-5}{kg/kg} \\
    $\tau_{s_\w}^{(L)}$ & \qty{2.04}{s} \\
    \hline\hline
  \end{tabular}
\end{table}

\section{Condensation rates in the statistical model}
\label{app:cr}
In this Appendix we derive the expressions for the conditional condensations rates $\langle C_\w \mid s_\w, t \rangle$ and $\langle C_\i \mid s_\w, t \rangle$, the mean condensation rates $\langle C_\w \rangle(t)$ and $\langle C_\i \rangle(t)$, and the statistical  model~\eqref{eq:SMC:s}. The notation $\langle {} \cdot {} \mid s_\w, t \rangle$ is a shorthand for the usual notation for the conditional averages:
\begin{equation}
    \langle {}\cdot{} \mid s_\w, t \rangle = \langle {} \cdot {} \mid S_\w(\bx, t) = s_\w \rangle.
\end{equation}
This defines the ensemble average over all flow realisations for which the supersaturation field at point $\bx$ at time $t$ equals $s_\w$. We use  upper- and lower-case letters to distinguish the random variable $S(\bx, t)$  from its value $s_\w$. All results here are valid for a statistically homogeneous system. For more details on the mathematical tools used here see Appendix~H in Ref.~\citep{pope2000turbulent}.

{\em Conditional condensation rates.} To simplify the notation, we ignore here the subscript $\w$ in $s_\w$ and use just $s$ for supersaturation, especially since the derivation is valid for $s_\i$ too. First we show that any conditional average for quantities of the form~\eqref{eq:condrates} can be computed~as
\begin{equation} \label{eq:basic_condensation_closure}
    \Bigl< \sum_{\alpha=1}^{N_\phi} G(\bx - \bx_\alpha) F\bigl(r_\alpha, S(\bx_\alpha,t) \bigr) \Bigm| S(\bx, t) = s \Bigr\rangle   
    =\frac{N_\phi}{V} \frac{f_\phi(s, t)}{f(s, t)} \, \langle F \mid s, t \rangle_\phi.
\end{equation}
Here $F$ is any function of a particle radius and supersaturation at the position of this particle, and $V$ is the volume of the simulation box. Next, $f(s, t)$ is the PDF of the Eulerian supersaturation field, it can be written as
\begin{equation}\label{eq:PDF_definition}
  f(s, t) = \langle \delta(S(\bx, t) - s) \rangle,
\end{equation}
where $\delta$ is the Dirac delta function.
The PDF $f_\phi$ of supersaturation at the particle positions (water droplets or ice particles, $\phi = \w$ or $\phi = \i$) is:
\begin{equation}
  f_\phi(s, t) = \frac{1}{N_\phi} \sum_{\alpha = 1}^{N_\phi} \langle \delta(S(\bx_\alpha, t) - s) \rangle.
\end{equation}
Averages over the particle positions are denoted as~$\langle {}\cdot {} \rangle_\phi$, $\phi = \w$ or $\phi = \i$:
\begin{equation}
  \langle F \mid s, t \rangle_\phi = \frac{1}{N_\phi} \sum_{\alpha = 1}^{N_\phi} \langle F(r_\alpha, S(\bx_\alpha,t)) \mid S(\bx_\alpha,t) = s \rangle.
\end{equation}
To derive~\eqref{eq:basic_condensation_closure}, we use that the conditional average of a field $H(\bx, t)$ can be expressed~as 
\begin{equation}\label{eq:the_formula}
  \bigl< H \bigm| S(\bx, t) = s \bigr> = \frac{1}{f(s, t)} \bigl< H \, \delta(S(\bx, t) - s) \bigr>.
\end{equation}
Second, we use that the system is statistically homogeneous. This allows us to  take the spatial average of the l.h.s.\ of Eq.~(\ref{eq:basic_condensation_closure}),
resulting in
\begin{align}
    \Bigl< & \sum_{\alpha=1}^{N_\phi} G(\bx - \bx_\alpha) F\bigl(r_\alpha, S(\bx_\alpha, t) \bigr) \Bigm| S(\bx, t) = s \Bigr>   
    = \sum_{\alpha=1}^{N_\phi} \frac{1}{f(s, t)} \bigl< G(\bx - \bx_\alpha) \, F\bigl(r_\alpha, S(\bx_\alpha, t) \bigr) \, \delta(S(\bx, t) - s) \bigr>  \\
    &= \frac{1}{V} \int_V \frac{1}{f(s, t)} \sum_{\alpha=1}^{N_\phi} \bigl< G(\bx - \bx_\alpha) \, F\bigl(r_\alpha, S(\bx_\alpha, t) \bigr) \, \delta(S(\bx, t) - s) \bigr> \, \upd \bx. \nonumber
    \end{align}
Third, we require that the support of spatial kernels $G$ is smaller than the length scale at which $s(\bx, t)$ varies, so that we can treat $G$ as a delta function. Under these conditions we obtain 
\begin{align}
        \Bigl\langle \sum_{\alpha=1}^{N_\phi} G(\bx - \bx_\alpha) & F\bigl(r_\alpha, S(\bx_\alpha, t\bigr) \Bigm| S(\bx, t) = s \Bigr\rangle 
        = \frac{1}{V} \frac{1}{f(s, t)} \sum_{\alpha=1}^{N_\phi} \bigl< F\bigl(r_\alpha, S(\bx_\alpha, t)\bigr) \, \delta(S(\bx_\alpha, t) - s) \bigr>  \\
        &= \frac{1}{V} \frac{f_\phi(s, t)}{f(s, t)} \sum_{\alpha=1}^{N_\phi} \langle F\bigl(r_\alpha, S(\bx_\alpha, t) \bigr) \mid S(\bx_\alpha, t) = s \rangle= \frac{N_\phi}{V} \frac{f_\phi(s, t)}{f(s, t)} \langle F \mid s, t \rangle_\phi.\nonumber 
    \end{align}
Using this result and Eq.~\eqref{eq:C_def}, we can rewrite the conditional condensation rates as
\begin{subequations} \label{eq:ccr}
    \begin{align}
      \langle C_\w \mid s_\w, t \rangle \label{eq:conditional_C_w}
      &= \frac{4}{3} \pi \frac{\rho_\w}{\rho_0} \frac{N_\w}{V} \frac{f_\w(s_\w, t)}{f(s_\w, t)} \biggl< \frac{\upd r_\w^3}{\upd t}  \biggm| s_\w, t \biggr>_{\!\w}, \\
      \langle C_\i \mid s_\w, t \rangle \label{eq:conditional_C_i}
      &= \frac{4}{3} \pi \frac{\rho_\i}{\rho_0} \frac{N_\i}{V} \frac{f_\i(s_\w, t)}{f(s_\w, t)} \biggl< \frac{\upd r_\i^3}{\upd t}  \biggm| s_\w, t \biggr>_{\!\i}.
    \end{align}
\end{subequations}
These formulae are consistent with Eq.~(5) in \citep{fries2024lagrangian}, who considered a turbulent mixing problem with spatially inhomogeneous initial conditions and strong phase change, using mapping-closure approximations.

{\em Mean condensation rates}.
Averaging expressions~(\ref{eq:ccr}) for $\langle C_\w \mid s_\w, t \rangle$ and $\langle C_\i \mid s_\w, t \rangle$ over $s_\w$, we  obtain Eqs.~\eqref{eq:Cav0} for the mean condensation rates.

{\em Statistical model for $s_\w'$ with condensation-rate fluctuations}. Now we derive the model~\eqref{eq:SMC:s} for the fluctuating supersaturation $s_\w'$ that involves the condensation terms following the method of~\cite{sarnitsky2022nonparametric}. We start with the exact equation for $s_\w'$, which for a statistically homogeneous system follows from~\eqref{eq:DNS:s_w}~as
\begin{equation}
    \frac{\upd s_\w'}{\upd t}
    = \varkappa \frac{\uppd^2 s_\w'}{\uppd x_j \uppd x_j} - A_{2, \w} C_\w' - A_{2, \i} C_\i' + f^{({s_\w})'}.
\end{equation}
If $s_\w'$ has a fast oscillating component we can model it as a stochastic differential equation
\begin{equation}\label{eq:general_SDE}
    \upd s_\w' = \Dodin(s_\w', t) \, \upd t + \sqrt{\Ddva(s_\w', t)} \, \upd W.
\end{equation}
This equation describes the behaviour of $s_\w'$ on timescales larger than $\tau_{\M, s_\w}$, the Markov--Einstein timescale of $s_\w$ and is not applicable for modelling the real behavior of $s_\w'$ on smaller timescales. The Markov--Einstein timescale is of the order of the Taylor timescale for $s_\w$, $\tau_{\M, s_\w} \sim \sqrt{\langle s_\w'^2 \rangle / \langle (\upd s_\w' / \upd t)^2 \rangle}$ and is of the order of the usual velocity Taylor timescale. 
In It\^o's stochastic calculus that we use in Eq.~\eqref{eq:general_SDE}, future values of white noise are independent on the current value of $s_\w'$ (the non-anticipation property). Since from~\eqref{eq:general_SDE} we can express the white-noise term~with
\begin{equation}
    \sqrt{\Ddva(s_\w', t)} \, \upd W = \upd s_\w' - \Dodin(s_\w', t) \, \upd t
\end{equation}
we requite it to obey the non-anticipation property:
\begin{equation}\label{eq:non-anticipation}
    \biggl< \frac{\upd s_\w'}{\upd t}(t + \Delta t) - \Dodin(s_\w(t + \Delta t), t + \Delta t) \biggm| s_\w, t\biggr> = 0, \quad\text{for $\Delta t \geq \tau_{M, s_\w}$}. 
\end{equation}
Since $\Dodin$ evolves on timescales larger than $\tau_{M, s_\w}$, we can approximate it from~\eqref{eq:non-anticipation}~as
\begin{equation}
    \begin{aligned}
        \Dodin(s_\w, t)
        &= \biggl< \frac{\upd s_\w'}{\upd t}(t + \tau_{M, s_\w}) \biggm| s_\w, t\biggr> = \varkappa \biggl< \frac{\uppd^2 s_\w'}{\uppd x_j \uppd x_j}(t + \tau_{M, s_\w}) \biggm| s_\w, t\biggr> + \bigl< f^{({s_\w})'}(t + \tau_{M, s_\w}) \mid s_\w \bigr> \\
        &\quad - A_{2, \w} \bigl< C_\w'(t + \tau_{M, s_\w}) | s_\w, t \bigr> - A_{2, \i} \bigl< C_\i'(t + \tau_{M, s_\w}) \mid s_\w, t \bigr>.
    \end{aligned}
\end{equation}
Because  the change of $s_\w$ due to evaporation/condensation is slow compared to $\tau_{\M, s_\w}$ in our setup, we can substitute $\langle C_\w'(t + \tau_{\M, s_\w}) \mid s_\w, t \rangle$ with $\langle C_\w'(t) \mid s_\w, t \rangle$. For the diffusion term, we can use a usual Langevin mixing closure, while the forcing term will be accounted with $\Ddva$ since its only role is to keep $\sigma_{s_\w}$ constant. Thus we turn~\eqref{eq:general_SDE} into
\begin{equation}
    \upd s'_\w = - A_{2,\w} \langle C_\w' \mid s_\w', t \rangle - A_{2,\i} \langle C_\i' \mid s_\w', t \rangle - \frac{1}{\tau_{s_\w}^{({L})}} s'_\w  \, \upd t + \sqrt{\Ddva(s'_\w, t}) \, \upd W(t).
\end{equation}
We could also relate $\Ddva(s_\w, t)$ to the statistics of $\upd s_\w' / \upd t$ as further described by~\cite{sarnitsky2022nonparametric}. However, for simplicity we just take $\Ddva$ to be independent of~$s_\w$, also ensuring that its value yields $\sigma_{s_\w} = \const$:
\begin{equation}\label{eq:Ddva}
    \Ddva = \frac{2 \sigma_{s_\w}^2}{\tau_{s_\w}^{({L})}} + 2 A_{2,\w} \langle C_\w' s_\w' \rangle + 2 A_{2,\i} \langle C_\i' s_\w' \rangle.
\end{equation}
Thus we arrive to Eq.~\eqref{eq:SMC:s}.

\section{Effect of phase change and small-scale 
turbulence on particle-size distributions}\label{app:moments}
In this Appendix we summarise details regarding the effect of small-scale turbulence on the mean supersaturation and the mean particle size, needed for the discussion in Section~\ref{sec:discussion}. For simplicity we neglect radius-dependent corrections, replacing Eq.~(\ref{eq:DNS:r}) by:
\begin{equation}\label{eq:r:simple}
    \frac{\upd r_\phi^2}{\upd t}
        = \biggl\{\begin{aligned}
            &2 A_{3, \phi} s_\phi &\text{if } r_\phi > 0, \\
            &0 &\text{if } r_\phi = 0.
        \end{aligned}
\end{equation}
Further, we consider the state when none of the particle considered has evaporated completely. Also, the derivations below rely on expanding $1/r_\phi$ around~$1/\langle r_\phi \rangle$ assuming small $r_\phi' / \langle r_\phi \rangle$,
\begin{equation}\label{eq:1/r}
    \frac{1}{r_\phi} = \frac{1}{\langle r_\phi \rangle} - \frac{r_\phi'}{\langle r_\phi \rangle^2} + \dots 
\end{equation}
and thus are valid for relatively sharp particle size distributions.

First we show that the correlation $\langle r'_\phi s'_\phi \rangle$ of the fluctuating quantities $r'_\phi $ and $s'_\phi $ is of the order of Da$_{r_\phi}$,
\begin{equation}\label{eq:rs_correlation}
    \langle r'_\phi s'_\phi \rangle = \frac{1}{2} \Da_{r_\phi} \frac{\sigma_{s_\phi}^2}{\langle s_\phi \rangle |\langle s_\phi \rangle|} \langle r_\phi \rangle \langle s_\phi \rangle,
\end{equation}
We start by deriving the evolution equation for $\langle r'_\phi s'_\phi \rangle$. From Eqs.~\eqref{eq:r:simple}, \eqref{eq:DNS:s_i}, \eqref{eq:SM:s} and~\eqref{eq:1/r}
we find 
\begin{equation}
    \begin{aligned}
        \frac{\upd \langle r'_\phi s'_\phi \rangle}{\upd t}
        &= \frac{\upd \langle r_\phi s'_\phi \rangle}{\upd t} = \biggl< \frac{\upd(r_\phi s'_\phi)}{\upd t} \biggr> = \biggl< r_\phi \frac{\upd s'_\phi}{\upd t} \biggr> + \biggl< s'_\phi \frac{\upd r_\phi}{\upd t} \biggr> \\
        &= -\frac{1}{\tau_{s_\w}^{(L)}} \langle r'_\phi s'_\phi \rangle + A_{3, \phi} \biggl< \frac{s_\phi s'_\phi}{r_\phi}\biggr> 
        = -\frac{1}{\tau_{s_\w}^{(L)}} \langle r'_\phi s'_\phi \rangle + A_{3, \phi} \frac{\sigma_{s_\phi}^2}{\langle r_\phi \rangle} + \ldots.
    \end{aligned}
\end{equation}
For small Damk\"ohler numbers, the evolution of the mean quantities, including $\langle r_\phi' s_\phi' \rangle$, happens on timescales much larger than the turbulent timescale $\tau_{s_\w}^{L}$. Hence we can neglect the term $\upd \langle r'_\phi s'_\phi \rangle / \upd t$ compared to the term $\langle r_\phi' s_\phi' \rangle / \tau_{s_\w}^{(L)}$. This results in 
\begin{equation}
    \langle r'_\phi s'_\phi \rangle
    = \tau_{s_\w}^{(L)} A_{3, \phi} \frac{\sigma_{s_\phi}^2}{\langle r_\phi \rangle}.
\end{equation}
Using the definitions~\eqref{eq:tau_r_phi} and~\eqref{eq:Da_r_phi} of $\tau_{r_\phi}$ and $\Da_{r_\phi}$ we arrive at Eq.~\eqref{eq:rs_correlation}.

Next, we derive~\eqref{eq:cwav_da}. The mean condensation rates are
\begin{equation} \label{eq:ccr:simple}
  \langle C_\phi \rangle
  = 4 \pi \frac{\rho_\phi}{\rho_0} A_{3, \phi} n_\phi \langle r_\phi s_\phi \rangle
  =  4 \pi \frac{\rho_\phi}{\rho_0} A_{3, \phi} n_\phi \bigl[ \langle r_\phi \rangle \langle s_\phi \rangle + \langle r_\phi' s_\phi' \rangle \bigr].
\end{equation}
Using~\eqref{eq:rs_correlation} we obtain~\eqref{eq:cwav_da}. The derivation of Eqs.~(\ref{eq:mre}) proceeds by averaging Eq.~\eqref{eq:DNS:r}, using Eq.~\eqref{eq:1/r} and Eq.~\eqref{eq:rs_correlation}:
\begin{align} \label{eq:pre_mean_r}
    \frac{\upd \langle r_\phi \rangle}{\upd t}
    &= A_{3, \phi} \biggl<\frac{s_\phi}{r_\phi} \biggr> = A_{3, \phi} \biggl<\frac{s_\phi}{\langle r_\phi \rangle} - \frac{s_\phi r_\phi'}{\langle r_\phi \rangle} \biggr>  \\
    &= A_{3, \phi} \frac{\langle s_\phi \rangle}{\langle r_\phi \rangle} \biggl( 1 - \frac{\langle r_\phi' s_\phi \rangle}{\langle r_\phi \rangle \langle s_\phi \rangle} \biggr) =  A_{3, \phi} \frac{\langle s_\phi \rangle}{\langle r_\phi \rangle} \biggr( 1 - \frac{1}{2} \frac{\sigma_{s_\phi}^2}{\langle s_\phi \rangle |\langle s_\phi \rangle|} \Da_{r_\phi} \! \biggr).  \nonumber
\end{align} 
\end{document}